\documentclass[aps,prb,twocolumn,superscriptaddress,floatfix,longbibliography]{revtex4-2}

\usepackage{amsmath,amssymb}    
\usepackage{bm}                 
\usepackage{graphicx}           
\usepackage{comment}            
\usepackage{braket}
\usepackage{physics}
\usepackage{tikz-cd}

\newcommand{\Ztwo}{\mathbb{Z}_2}
\newcommand{\dotprod}[2]{#1\cdot#2}

\newcommand{\chainaccent}[2][c]{\mathbf{#1}_{#2}'}

\newcommand{\chain}[2][c]{\mathbf{#1}_{#2}}
\newcommand{\bound}[1]{\partial_{#1}}
\newcommand{\cycle}[2][c]{\bound{#2}\chain[#1]{#2}}

\newcommand{\dualchain}[2][c]{\mathbf{\overline{#1}}_{#2}}
\newcommand{\dualbound}[1]{\overline{\partial}_{#1}}
\newcommand{\dualcycle}[2][c]{\dualbound{#2}\dualchain[#1]{#2}}

\newcommand{\cell}[1][k]{\chain[q]{#1}}
\newcommand{\face}[1][k]{\chain[f]{#1}}
\newcommand{\edge}[1][k]{\chain[e]{#1}}
\newcommand{\vertex}[1][k]{\chain[v]{#1}}

\newcommand{\dualcell}[1][k]{\dualchain[q]{#1}}
\newcommand{\dualface}[1][k]{\dualchain[f]{#1}}
\newcommand{\dualedge}[1][k]{\dualchain[e]{#1}}
\newcommand{\dualvertex}[1][k]{\dualchain[v]{#1}}

\newcommand{\miller}[1]{\left[#1\right]}
\newcommand{\qbound}[2]{\bound{#1}^{\miller{#2}}}

\newcommand{\Miller}[1]{\left[\mathbf{#1}\right]}
\newcommand{\Qbound}[2]{\bound{#1}^{\Miller{#2}}}
\newcommand{\Qdualbound}[2]{\dualbound{#1}^{\Miller{#2}}}

\newcommand{\plusket}{\Ket{+}}

\newcommand{\GHZ}[1][]{\Ket{\text{GHZ}_{#1}}}

\newcommand{\fatm}{\mathbf{m}}

\begin{document}

\title{Fault-tolerant structures for measurement-based quantum computation on a network}

\author{Yves van Montfort}
\affiliation{QuTech, Delft University of Technology, Lorentzweg 1, 2628 CJ Delft, The Netherlands}
\author{Sébastian de Bone}
\affiliation{QuTech, Delft University of Technology, Lorentzweg 1, 2628 CJ Delft, The Netherlands}
\affiliation{QuSoft, CWI, Science Park 123, 1098 XG Amsterdam, The Netherlands}
\author{David Elkouss}
\email[]{david.elkouss@oist.jp}
\affiliation{QuTech, Delft University of Technology, Lorentzweg 1, 2628 CJ Delft, The Netherlands}
\affiliation{Networked Quantum Devices Unit, Okinawa Institute of Science and Technology Graduate University, Okinawa, Japan}

\date{\today}

\begin{abstract}
In this work, we introduce a method to construct fault-tolerant \emph{measurement-based quantum computation} (MBQC) architectures and numerically estimate their performance over various types of networks. A possible application of such a paradigm is distributed quantum computation, where separate computing nodes work together on a fault-tolerant computation through entanglement. We gauge error thresholds of the architectures with an efficient stabilizer simulator to investigate the resilience against both circuit-level and network noise. We show that, for both monolithic (\textit{i.e.}, non-distributed) and distributed implementations, an architecture based on the diamond lattice may outperform the conventional cubic lattice. 
Moreover, the high \emph{erasure} thresholds of non-cubic lattices may be exploited further in a distributed context, as their performance may be boosted through \emph{entanglement distillation} by trading in entanglement success rates against erasure errors during the error-decoding process. 
These results highlight the significance of lattice geometry in the design of fault-tolerant measurement-based quantum computing on a network, emphasizing the potential for constructing robust and scalable distributed quantum computers.
\end{abstract}

\maketitle

\section{Introduction}\label{sec:introduction}
Large-scale quantum computation with low error rates requires handling noise in a correct and efficient manner---one would like to \emph{fault-tolerantly} transmit, store, and process quantum information with quantum computing hardware. Quantum error-correction encompasses methods to achieve fault tolerance from faulty hardware. Topological error-correction codes, including surface codes, are a promising avenue to achieve this goal, as these codes have high error thresholds against local errors, and only require nearest-neighbor interactions between qubits in a two-dimensional layout.

Surface codes achieve fault-tolerance by repeatedly combining measurement outcomes of consecutive rounds of stabilizer measurements~\citep{fujiiQuantumComputationTopological2015}. Geometrically, the addition of a time dimension to a two-dimensional decoding (\textit{i.e.}, \emph{syndrome}) graph creates a three-dimensional structure, that may be interpreted as a noisy quantum channel where logical information is propagated in the direction of time. For example, the conventional planar or toric surface code may be constructed from qubits on a square lattice, but its corresponding decoding graphs are three-dimensional cubic lattices. 

\emph{Foliation}~\cite{boltFoliatedQuantumErrorCorrecting2016} is a method to transform any surface code to a three-dimensional \emph{cluster state} that forms the resource for fault-tolerant \emph{measurement-based quantum computation} (MBQC), with the same geometry as the corresponding logical quantum channel of the surface code~\cite{raussendorfFaulttolerantOnewayQuantum2006, barrettFaultTolerantQuantum2010}. Although foliated codes can be interpreted as having replaced time with another spatial dimension, the resulting cluster states can be initialized and consumed in arbitrary directions over time, lifting the rigid duality of time and space of surface codes. A concrete example is so-called \emph{interleaving} in fusion-based quantum computation~\cite{bombinInterleavingModularArchitectures2021}, which bears many resemblances to the measurement-based architectures considered here.

However, foliation does not exhaust all possible fault-tolerant cluster states---\textit{i.e.}, there are \emph{non-foliated} three-dimensional cluster states that cannot be constructed from a two-dimensional surface code. We study these types of lattices because they may produce higher fault-tolerant error thresholds than conventional surface codes, at least when assuming simple combinations of independent and identically distributed (i.i.d.) single-qubit and measurement errors~\cite{nickersonMeasurementBasedFault2018}. In practice though, non-foliated lattices will not necessarily produce higher error thresholds, because faulty operations during cluster state preparation and measurement typically introduce highly coupled and non-identically distributed errors. Both the complexity of the error decoder (through the syndrome graph) and the complexity of the error model (through the quantum circuit) eventually determine the error threshold.

Because state-of-the-art hardware is currently not capable of realizing large cluster states that can achieve sufficiently low logical error rates, some research has focused its attention to modular implementations of fault-tolerant MBQC states~\cite{augerFaulttolerantQuantumComputation2018, herrLocalScalableLattice2018, bartolucciFusionbasedQuantumComputation2023}. The entire cluster state is prepared from resource states that are generated by small, separate devices, and entangling operations between devices create the entanglement to encode the entire cluster state. Several physical systems are suitable for modular MBQC architectures via optical interfaces~\cite{awschalomQuantumTechnologiesOptically2018}.

In this paper, we explore fault-tolerant cluster states under more realistic noise models compared to previous investigations~\citep{nickersonMeasurementBasedFault2018}. This brings their realization in MBQC closer to reality. We numerically estimate fault-tolerant thresholds of previously proposed cluster states~\citep{nickersonMeasurementBasedFault2018} for both \emph{monolithic} (\textit{i.e.}, non-distributed) and distributed implementations. We use an efficient stabilizer simulator and the Union-Find decoder~\citep{delfosseAlmostlinearTimeDecoding2021} to evaluate the performance of our architectures. In our distributed noise models, we consider circuit-level noise during state preparation, measurement noise of single-qubit measurements, and \emph{network noise} between nodes introduced by qubits prepared in the Greenberger-Horne-Zeilinger (GHZ) basis. 

We investigate the first step towards including entanglement distillation for one of the distributed cluster states considered and find that including distillation is particularly effective in the context of measurement-based fault tolerance. This is due to the high \emph{erasure}-type error thresholds of non-foliated lattices that may shield against probabilistic entanglement generation resulting from distillation.

\section{Cluster state construction}\label{sec:code_construction}
In this section, we briefly summarize the mathematical description of cluster states used throughout this paper along the same lines as Fujii~\cite{fujiiQuantumComputationTopological2015}. We assume that the reader is already familiar with definitions of the Pauli group, the Clifford group and the stabilizer group, which may be found in detail in Nielsen and Chuang~\cite{Nielsen2000}.

In Sec.~\ref{subsec:chain_complex}, we introduce relevant concepts of a $\Ztwo$ chain complex. In Sec.~\ref{subsec:three_dimensional_cluster_codes}, we define a fault-tolerant three-dimensional cluster state using such a chain complex. In Sec.~\ref{subsec:cluster_error_correction}, we describe the error correction process for three-dimensional cluster states. 

\subsection{$\Ztwo$ chain complex}\label{subsec:chain_complex}
A $\Ztwo$ chain complex starts with a definition of \emph{vector spaces} $C_i$, each constructed over the field $\Ztwo$ and indexed with the \emph{dimension} $i\in\{0,1,\dots,D\}$. These vector spaces form a sequence $C_D \rightarrow C_{D-1} \rightarrow \dots \rightarrow C_0$, where subsequent pairs are connected by homomorphisms $\bound{i}: C_i \rightarrow C_{i-1}$ called \emph{boundary operators}. 

In this work, we construct chain complexes over a three-dimensional set $\mathcal{S}=\left(Q,F,E,V\right)$ of cells (\textit{i.e.} volumes) $Q=\{\cell\}$, faces $F=\{\face\}$, edges $E=\{\edge\}$ and vertices $V=\{\vertex\}$. Cells, faces, edges and vertices form the basis elements of the vector spaces $C_3$, $C_2$, $C_1$ and $C_0$, respectively:
\begin{center}
\begin{tikzcd}[row sep = tiny]
C_3 \arrow{r}{\bound{3}} & C_2 \arrow{r}{\bound{2}} & C_1 \arrow{r}{\bound{1}} & C_0 \\
\{\cell\} & \{\face\} & \{\edge\} & \{\vertex\}
\end{tikzcd}
\end{center}
An element of $C_i$ is called an $i$-chain and notated as $\chain{i}$. Such a chain can be considered a linear combination of basis elements of $C_i$ with $\Ztwo$ coefficients. For example, a 1-chain $\chain{1}\in C_1$ is a combination of edges:
\begin{equation}
\label{eq:chain-def}
    \chain{1} = \sum_k z_k \edge \equiv \begin{bmatrix} z_0 & z_1 & \cdots \end{bmatrix}^T \quad \text{where} \quad z_k \in \Ztwo.
\end{equation}
Here, we take the vector representation with respect to the basis $E=\{\edge\}$. Analogous definitions hold for the remaining vector spaces and their basis elements.

Boundary operators $\bound{i}:C_i \rightarrow C_{i-1}$ are linear operators:
\begin{equation}
    \bound{i}\left(\chain{i} + \chainaccent{i}\right) = \cycle{i} + \bound{i}\chainaccent{i}.
\end{equation}
For chain complexes defined on $\mathcal{S}$, boundary operators have an intuitive geometric interpretation. A cell $\cell$ is mapped to the faces $\{\face[j]\}$ that enclose it, a face $\face$ is mapped to the edges $\{\edge[j]\}$ that form its boundary and an edge $\edge$ is mapped to its endpoints $\{\vertex[j]\}$, which will always be a pair of vertices on the graph $\left(E,V\right)$. By definition, two boundary maps applied in succession on a chain $\chain{i}$ produce the zero map
\begin{equation}
\label{eq:doublebound}
    \bound{i-1}\bound{i} = 0,
\end{equation}
no matter the choice of $\chain{i}$. We equip vector spaces $C_i$ with the standard inner product
\begin{equation}
\label{eq:chaininner}
    \dotprod{\chain{i}}{\chainaccent{i}} \equiv \chain{i}^T \chainaccent{i},
\end{equation}
that is, a dot product with addition modulo 2 over the pairwise multiplied coefficients. The inner product may be geometrically interpreted as a basis-independent parity measurement of the ``overlap'' of two $i$-chains.
An $i$-chain $\chain{i}$ with a zero boundary (\textit{i.e.}, $\cycle{i}=0$) is called an $i$-\emph{cycle} (or simply, \emph{cycle}). Note that $i$-cycles form a group under element-wise addition. A cycle $\chain{i}$ is called \emph{trivial} whenever there exists some chain $\chain{i+1}\in C_{i+1}$, such that $\chain{i}=\cycle{i+1}$. Cycles that cannot be formed in this way are called \emph{non-trivial}. Trivial cycles form a normal subgroup of all cycles, such that the quotient groups
\begin{equation}
\label{eq:homology}
    H_i \equiv \ker\bound{i} / \operatorname{Im} \bound{i+1}
\end{equation}
divide cycles into equivalence classes that are trivially related. The groups $H_i$ are called \emph{homology groups}, where two chains $\chain{i}$ and $\chainaccent{i}$ belong to the same class whenever $\chain{i}=\chainaccent{i}+\cycle{i+1}$.

The \emph{dual complex} is another sequence $\overline{C}_D \rightarrow \overline{C}_{D-1} \rightarrow \dots \rightarrow \overline{C}_0$, where each $\overline{C}_i$ shares the structure of $C_{D-i}$. Associated with the dual complex are the dual boundaries $\dualbound{i}: \overline{C}_i \rightarrow \overline{C}_{i-1}$. Chains in the dual complex are called \emph{dual chains} or \emph{co}chains, whereas those referring to the original complex are \emph{primal} chains or, more succinctly, chains. For the three-dimensional lattice $\mathcal{S}$, vertices map to dual cells $\left(V \rightarrow \overline{Q} \right)$, edges map to dual faces $\left(E \rightarrow \overline{F} \right)$, faces map to dual edges $\left(F \rightarrow \overline{E} \right)$, and cells map to dual vertices $\left(Q \rightarrow \overline{V} \right)$, such that:
\begin{center}
\begin{tikzcd}[row sep = tiny]
\overline{C}_0 & \arrow{l}{\dualbound{1}} \overline{C}_1 & \arrow{l}{\dualbound{2}} \overline{C}_2 & \arrow{l}{\dualbound{3}} \overline{C}_3 \\
\{\dualvertex\} & \{\dualedge\} & \{\dualface\} & \{\dualcell\}
\end{tikzcd}
\end{center}
The dual boundaries on $\overline{\mathcal{S}}=\left(V,E,F,Q\right)$ behave similarly to their primal counterparts. With a slight abuse of notation, a dual cell $\dualcell=\vertex$ is mapped to the faces $\{\dualface[j]\}$ that enclose it, which correspond to the primal edges $\{\edge[j]\}$ \emph{incident} to $\vertex$. A dual edge $\dualedge=\face$ is mapped to the endpoints $\{\dualvertex[j]\}$, which are the cells $\{\cell[j]\}$ adjacent to $\face$. Similar to cycles, $i$-\emph{cocycles} are dual chains $\dualchain{i}$ that have no coboundary. Dual boundaries share a similar concept as homology, called \emph{cohomology}. Cohomology groups are formed in the same way as homology groups (Eq. \ref{eq:homology}), by replacing cycles $\ker\bound{i}$ and trivial cycles $\operatorname{Im} \bound{i+1}$ by the dualized versions $\ker\dualbound{i}$ and $\operatorname{Im} \dualbound{i+1}$, such that
\begin{equation}
   \label{eq:cohomology}
   \overline{H}_i \equiv \ker\dualbound{i} / \operatorname{Im} \dualbound{i+1}.
\end{equation}

\subsection{Three-dimensional cluster states}\label{subsec:three_dimensional_cluster_codes}
We now make use of chain complexes $C_3 \rightarrow C_2 \rightarrow C_1 \rightarrow C_0$ to construct a fault-tolerant cluster state on a three-dimensional lattice $\mathcal{S}$, given the following recipe:
\begin{enumerate}
    \item Place qubits on all basis elements of $C_2$ and $\overline{C}_2$, \textit{i.e.}, on all faces $\face$ and edges $\edge$ of the lattice. Pauli operators on the qubits of an $i$-chain $\chain{i}$ are denoted as $\sigma(\chain{i})\equiv\prod_k \sigma^{z_k}$, where $\sigma\in\{X,Y,Z\}$. Qubits with $z_k=0$ carry identity, and $z_k=1$ carry $\sigma$.
    \item For each face $\face$, define a \emph{primal} stabilizer generator $g_k=X(\face)Z(\bound{2}\face)$. These operators generate stabilizers $X(\chain{2})Z(\cycle{2})$ for all 2-chains $\chain{2}\in C_{2}$.
    \item For each edge $\edge$, define a \emph{dual} stabilizer $\overline{g}_k=X(\edge)Z(\dualbound{2}\edge)$. These operators generate stabilizers $X(\dualchain{2})Z(\dualcycle{2})$ for all 2-cochains $\dualchain{2}\in \overline{C}_2$.
\end{enumerate}
\begin{figure}[t]
    \centering
    \includegraphics[width=\linewidth]{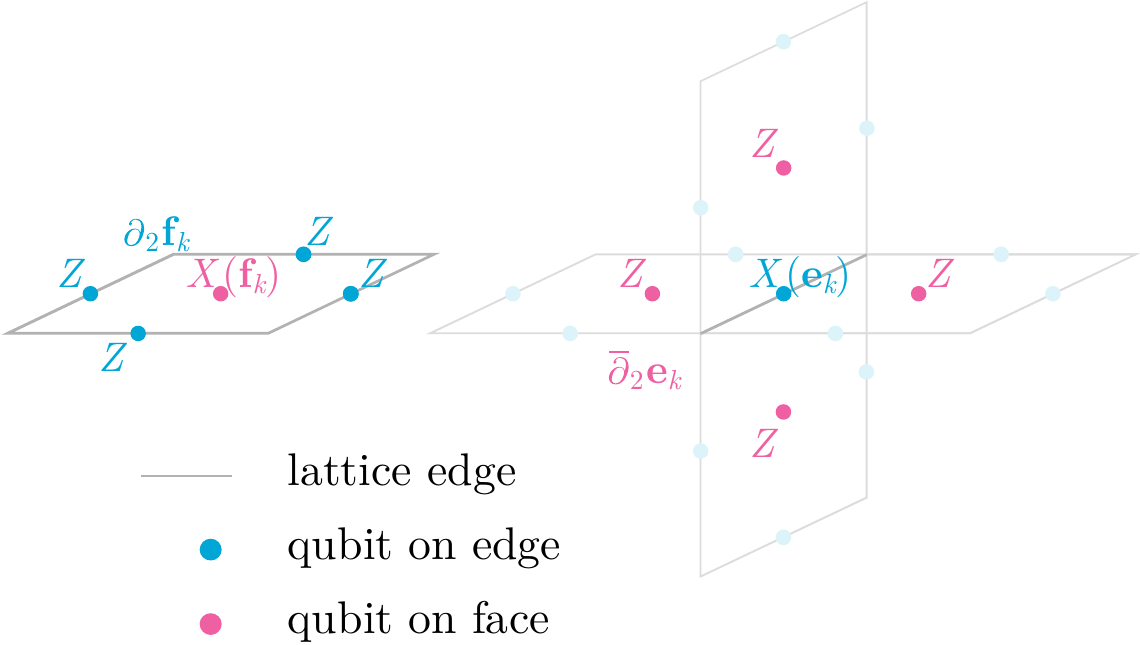}
    \caption[Stabilizers of a three-dimensional cluster state.]{A primal and dual stabilizer generator for a three-dimensional cluster state as defined in Sec.~\ref{subsec:three_dimensional_cluster_codes}. The primal generator $X(\face)Z(\bound{2}\face)$ is associated with the highlighted face on the left of the figure. The dual generator $X(\edge)Z(\dualbound{2}\edge)$ is associated with the highlighted edge on the right of the figure. Stabilizer generators associated with the other faces and edges are not shown explicitly.}
    \label{fig:stabilizers_cluster_state}
\end{figure}
We depict an example of a primal and dual stabilizer generator associated with a single face $\face$ and edge $\edge$ in Fig.~\ref{fig:stabilizers_cluster_state}. Note that each stabilizer generator carries a Pauli-$X$ operator on some face (edge) qubit and Pauli-$Z$ operators on its direct neighbors on the (dual) boundary. This stabilizer composition is commonly associated with \emph{cluster} or \emph{graph} states~\cite{briegelPersistentEntanglementArrays2001}. Logical operators are derived from elements of the (co)homology groups of the underlying chain complex. For the logical identity channel, logical $X$-type operators correspond with elements of the homology group $H_1$, and logical $Z$-type operators correspond with elements of the cohomology group $\overline{H}_1$. The construction of a fault-tolerant channel from a cluster state that carries logical information is subtle: for details we refer the reader to Ref.~\citep{fujiiQuantumComputationTopological2015}. We note that logical qubits are usually introduced by creating lattice boundaries---such as holes inside the bulk---by switching off stabilizers on these boundaries.

\subsection{Error correction}\label{subsec:cluster_error_correction}
For the cluster states described above, we can construct the primal (dual) \emph{error syndrome} by measuring out the qubits on the faces (edges) in the Pauli-$X$ basis. For each face $\face$ (edge $\edge$), this leads to a measurement outcome $\mu_k\in\{+1,-1\}$ ($\overline{\mu}_k\in\{+1,-1\}$). 
Error syndromes are constructed from measurement outcomes in the following way:
\begin{enumerate}
    \item For each cell $\cell$, we produce a \emph{primal} error syndrome $m_k$ as the product of measurement outcomes of the qubits that lie on its boundary: $m_k=\prod_{\face[j]\in\bound{3}\cell}\mu_j$. Note that $m_k$ represents the measurement outcome of the stabilizer $s_k\equiv\prod_{\face[j]\in\bound{3}\cell}g_{j}=X(\bound{3}\cell)Z(\bound{2}\bound{3}\cell)=X(\bound{3}\cell)$.
    \item For each dual cell $\dualcell$ (i.e., vertex $\vertex$), we produce a \emph{dual} error syndrome $\overline{m}_k=\prod_{\edge[j]\in\dualbound{3}\vertex}\overline{\mu}_j$. The syndrome represents the measurement outcome of the stabilizer $\overline{s}_k\equiv\prod_{\edge[j]\in\dualbound{3}\vertex}\overline{g}_j=X(\dualbound{3}\vertex)$. 
\end{enumerate}

Note that in the absence of errors, all error syndromes produce outcomes $m_k=+1$ and $\overline{m}_k=+1$. Since all cluster state qubits are measured out in the Pauli-$X$ basis, we can restrict ourselves to probabilistic Pauli-$Z$ errors---\textit{i.e.}, phase-flips---on the qubits prior to syndrome measurement. These errors can appear \emph{phenomenologically} as a result of i.i.d.~sampling, or for example as a result of depolarizing or dephasing noise after a circuit or network operation.  
This model also includes measurement errors, because such an error is equivalent to a probabilistic $Z$ gate before an $X$ basis measurement. Measurement outcomes are then fully described by (anti)commutation relations of Pauli operators.

For notational convenience, let a particular set of primal error syndromes be described by a vector $\mathbf{m}$ over the field $\Ztwo$, where the $k$th element $m_k$ maps measurement outcomes $\{+1,-1\}\mapsto\{0, 1\}$. Given a set of Pauli-$Z$ errors on dual edges (primal faces) described by a dual chain $Z(\dualchain{1})$, the entire error syndrome takes the convenient form
\begin{equation}
\label{eq:errorsyndrome}
    \mathbf{m}=\dualbound{1}\dualchain{1}.
\end{equation}
That is, the primal error syndrome corresponds to the \emph{boundary} of all the $Z$-type errors on face qubits $\face$. Similarly, the dual syndrome $\overline{\mathbf{m}}=\bound{1}\chain{1}$ is the boundary of Pauli-$Z$ errors described by the chain $Z(\chain{1})$.

The decoding problem may now be stated as follows. Given a pair of primal and dual syndrome outcomes $\{\mathbf{m},\overline{\mathbf{m}}\}$, we identify recovery chains $\dualchain[r]{1}$ and $\chain[r]{1}$ such that $\dualbound{1}(\dualchain[r]{1}+\dualchain{1})=0$ and $\bound{1}(\chain[r]{1}+\chain{1})=0$---\textit{i.e.}, the sum of recovery and error chains form \emph{cycles} in the corresponding chain complex. We identify a \emph{logical failure} whenever decoding introduces a logical $X$-type and/or $Z$-type error across the channel, which occurs whenever the sum of a recovery and error chain forms a \emph{non-trivial} cycle.

We emphasize that we only evaluate fault tolerance of the cluster state as a pure \emph{quantum memory}. That is, the noise thresholds that we determine exclusively assess the state's ability to protect logical information and do not, \textit{e.g.}, include the operations required to encode this logical information or operate on it.

\section{Methods}\label{sec:methods}
\begin{figure}[t]
    \centering
    \includegraphics[width=0.95749674796\linewidth]{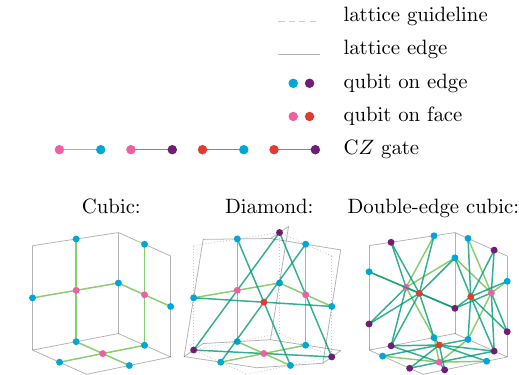}
    \caption[Cluster states obtained through splitting.]{Cluster states obtained through splitting.
    The cubic cluster state can be created by foliating the standard toric surface code. The diamond cluster state is obtained through two splits of the primal and dual vertex in the cubic unit cell. The complex is regular and self-dual, with each face connected to six edges. The double-edge cubic cluster state is obtained through multiple simple splits of primal and dual vertices. Each face is ``double''-sided, supporting two different qubit that are connected to eight surrounding edges.}
    \label{fig:naomi-clusters}
\end{figure}

In the current work, we are interested in crystalline cluster states that are built from cellulations of flat three-dimensional space. There are various methods to find such structures. Two approaches that have previously been used are the splitting procedures on a known (foliated) structure, such as the cubic cluster state~\cite{nickersonMeasurementBasedFault2018}, and an algebraic approach based on combinatorial tiling theory~\cite{newmanGeneratingFaultTolerantCluster2020}. We briefly discuss the former method below and show some of the lattices that were found through this method in Fig.~\ref{fig:naomi-clusters}. Although we have not used the latter method to construct new cluster states, we emphasize that the zoo of fault-tolerant cluster states merits further investigation under noise models considered here.

Below, we first discuss an extension of the concepts of the chain complex as discussed in Sec.~\ref{subsec:chain_complex}. By adding indices that describe translational symmetry, we can use the chain complex to describe unit cells of a lattice that generate a full lattice. We discuss this method in Sec.~\ref{subsec:unit_cell_and_crystal}. In Sec.~\ref{subsec:splitting-1}, we describe the \emph{cell-vertex splitting} operation of Nickerson and Bombín~\cite{nickersonMeasurementBasedFault2018} in the context of this unit cell complex---this allows us to transform unit cells that describe three-dimensional lattices. In Sec.~\ref{sec:splitting-2}, we define a \emph{face-edge splitting} operation that allows us to replace cluster state qubits with entangled states or Bell measurements. In Sec.~\ref{sec:entanglement}, we introduce the noise models used for monolithic (\textit{i.e.}, circuit-level) and network noise, and describe our method for generating entanglement in a distributed cluster state. In Sec.~\ref{subsec:numerical_tools_and_considerations}, we discuss how we use the stabilizer formalism to model and transfer Pauli errors in the circuits that we use to construct and measure cluster states, and we elaborate on the numerical aspects of our model and simulations.

\subsection{Unit cell complex}\label{subsec:unit_cell_and_crystal}
The \emph{unit cell complex} is the set of basis elements (atoms) together with their boundary relations (bonds) in the crystal that forms a block with translation symmetry along the sides of the unit cell. We use the \emph{Miller index} notation with square brackets $\miller{abc}$ or (slightly unorthodox yet succinct) $\Miller{r}$ for a translation $\mathbf{r}\equiv a\mathbf{x} + b\mathbf{y} + c\mathbf{z}$ along the \emph{lattice vectors} $\mathbf{x}$, $\mathbf{y}$ and $\mathbf{z}$ that form the sides of the unit cell. Negative indices are denoted in the usual way as $\miller{\overline{abc}}$ or $\Miller{\overline{r}}$ for a translation $\mathbf{\overline{r}} \equiv -a\mathbf{x} - b\mathbf{y} - c\mathbf{z}$.

The construction of the unit cell complex follows from the choice of lattice vectors. The subset of basis elements $\{\chain[b]{i}\}$ in $C_i$ that are equivalent under translations $\mathbf{r}$ is mapped to a single \emph{quotient} element $\chain[q]{i}$, which serves as a basis vector for a new vector space $Q_i$ over $\Ztwo$. Similarly, the boundary relation between elements $\left(\chain[b]{i}\right)_n$ and $\left(\chain[b]{i-1}\right)_m$ is mapped to a relation to their quotient elements $\left(\chain[q]{i}\right)_n$ and $\left(\chain[q]{i-1}\right)_m$ in the form of a \emph{quotient} boundary map $\Qbound{i}{r}: Q_i \mapsto Q_{i-1}$. Because some of the boundary relations are present between elements \emph{across} two unit cells (such as a face with boundary edges from adjacent unit cells), quotient boundaries have a Miller index $\Miller{r}$ that represents the translation $\mathbf{r}$ required to jump to its neighboring unit cell. Intracellular boundaries are encoded by $\qbound{i}{0}$, whilst intercellular boundaries are encoded by $\qbound{i}{abc}$ for a non-zero translation $\mathbf{r}\equiv a\mathbf{x} + b\mathbf{y} + c\mathbf{z}$. A more detailed description of the quotient boundaries in the unit cell complex can be found in App.~\ref{subsec:unit_cell_complex}.

Given a unit cell complex, an \emph{embedding} is a map that takes each $Q_i$ and the quotient maps $\Qbound{i}{r}$ to a crystalline chain complex $C_3\rightarrow C_2 \rightarrow C_1 \rightarrow C_0$, with lattice dimensions given by the embedding. For a periodic lattice of $N \equiv N_\mathbf{x} \times N_\mathbf{y} \times N_\mathbf{z}$ unit cells along the $\mathbf{x}$, $\mathbf{y}$, and $\mathbf{z}$ lattice directions, respectively, vector spaces of chains take the form
\begin{equation}
    C_i = Q_i^{\oplus N} = Q_i \otimes L.
\end{equation}
Here, $L \equiv \Ztwo^{\oplus N}$ is an $N$-dimensional vector space over $\Ztwo$. The $N$ basis vectors of $L$ represent $N$ lattice points, such that linear combinations in $L$ may be associated with the subset of lattice points that have non-zero coefficients. Intuitively, the embedding is realized by repeating the unit cell elements $Q_i$ over each lattice point given by a displacement vector $\mathbf{r}$. We formalize this process in App.~\ref{subsec:crystal_embeddings}.

\subsection{Cell-vertex splitting}\label{subsec:splitting-1}
The splitting procedure of Nickerson and Bombín~\cite{nickersonMeasurementBasedFault2018} may be phrased in terms of the unit cell complex introduced in Sec.~\ref{subsec:unit_cell_and_crystal}. A visual example of a split in two dimensions can be seen in App.~\ref{subsec:crystal_embeddings}, Fig.~\ref{fig:unit-cells}a and b; splitting each face of a square lattice diagonally results in the triangular lattice shown adjacently. Here, we review the general case of an $n$-split in three dimensions, with a simple split following from the case $n=1$. Importantly, the splitting number $n$ alone is not sufficient to uniquely characterize a split: one should also specify the new boundary relations between split vertices and edges. Denote the split vertex with $\vertex[0]$. The recipe for an $n$-split is then as follows:

\begin{enumerate}
    \item Let $E' = \set{(\vertex[j], \vertex[0]) | \vertex[j] \in N_0}$ be the set of incident edges on $\vertex[0]$ with neighborhood $N_0$. Choose $n$ disjoint subsets $E_i' \in E'$ ($i=1 \dots n$) that will each connect to a new vertex.
    \item Create $n$ new vertices $\vertex[i]$, and connect each $\vertex[i]$ to the incident edges $E_i'$. The $\vertex[0]$ vertex connects to the remaining edges $E_0' = E' \setminus \bigcup_i E_i'$, which might be the empty set.
    \item Create $n$ new edges $\edge[i] = (\vertex[i], \vertex[0])$. The corresponding boundary relations are encoded in $\qbound{1}{0}$ of the unit cell complex with Miller index $[0]$.
    \item Fix the remaining boundary maps $\Qbound{2}{r}$. That is, for each $\vertex[i]$ and $\forall\mathbf{r}$, calculate the dual boundary $\chain{2} \equiv \sum_{\mathbf{p}}\Qdualbound{2}{p}\Qdualbound{3}{r-p}\;\vertex[i]$. By the zero map conditions (Eq.~\eqref{eq:doubleqbound} in App.~\ref{subsec:unit_cell_complex}), the right-hand side should be zero. That is, we connect faces $\face[j] \in \chain{2}$ to the newly created edge $\edge[i]$ with Miller index $\mathbf{\overline{r}}$.
\end{enumerate}

\subsection{Face-edge splitting}
\label{sec:splitting-2}
The cell-vertex splitting procedure described before in Sec.~\ref{subsec:splitting-1} changes both the number of syndromes and the connectivity between syndromes in the syndrome graph. We can define an additional splitting operation on faces (dual edges) of such a complex. We discuss this operation in this section.

Usually, a cluster state as described in Sec.~\ref{subsec:three_dimensional_cluster_codes} is constructed with the aid of C$Z$ gates on qubits initialized in the $\ket{+}$ state. Alternatively, one may replace C$Z$ gates with other entangling operations that lead to the same stabilizer states. 

\begin{figure}[t]
    \centering
    \includegraphics[width=\linewidth]{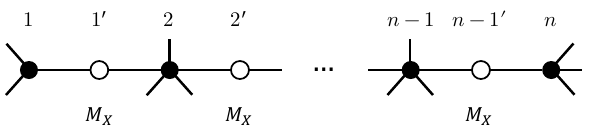}
    \caption[Bell/GHZ state subgraph.]{A subgraph of $2n-1$ qubits. Qubits marked primed indices are measured in the $X$ basis. Unmeasured qubits are connected arbitrarily to the rest of the graph. If one assumes that every measurement outcome $m=0$, the unmeasured qubits can be initialized in an $n$-partite Bell/GHZ state.}
    \label{fig:alternating-ghz}
\end{figure}

Consider the subgraph of a graph state as in Fig.~\ref{fig:alternating-ghz}. Qubits at odd positions are marked with integers $i\in\{1,2,\dots,n\}$ and their right 
neighbors at even positions with a primed index $i' \neq n'$. In this graph state, odd qubits have an arbitrary number of neighbors, whereas even qubits only neighbor the two odd qubits on either side. 
After measuring the even qubits $i'$ in the Pauli-$X$ basis, the post-measurement $m_{i'}=0$ graph state stabilizers are
\begin{equation}
\begin{split}\label{eq:post_measurement_subgraph_stabilizers}
S_{\text{post}} = \langle &X_{1'},X_{2'},\dots,X_{n-1'}, \\
&\prod_{i=1}^n(X_i\prod_{j\in N(i)}Z_j), \\
&Z_1 Z_2, Z_2 Z_3, \dots, Z_{n-1}Z_n\rangle.
\end{split}
\end{equation}
Here, $N(i)=\set{j|(i,j)\in E}$ are the qubits connected to the odd qubit $i$ \emph{outside} the subgraph depicted in Fig.~\ref{fig:alternating-ghz}. 
The disentangled measured qubits play no further role in the graph state stabilizers. In practice, one may replace these ``virtual'' qubits with measurement outcomes $m=0$ and initialize the unmeasured qubits in the state stabilized by Eq.~\eqref{eq:post_measurement_subgraph_stabilizers}---the resulting state is the same.
In constructing graph states, C$Z$ gates transform an $X$-type stabilizer $\prod_{i=1}^n X_i \mapsto \prod_{i=1}^n (X_i\prod_{j \in N(i)}Z_j)$, whilst leaving the $Z$-type stabilizers untouched. This means that we can alternatively initialize the odd qubits $i$ as an $n$-qubit $\GHZ[n]$ state stabilized by $\langle \prod_{i=1}^n X_i, Z_1 Z_2, Z_2 Z_3, \dots, Z_{n-1}Z_n\rangle$, before applying the C$Z$ gates to qubits outside the subgraph. If $n=2$, one can initialize with the bipartite Bell state $\Braket{X_1X_2, Z_1Z_2}$. 

The above procedure shows how the subgraph in Fig.~\ref{fig:alternating-ghz} need not be initialized through C$Z$ gates, so long as there is a protocol that can create $\GHZ[n]$. Our new splitting procedure on faces produces subgraphs like Fig.~\ref{fig:alternating-ghz} in a systematic way. Importantly, the closed-cell stabilizers of the cluster state stay intact on the split geometry.

\begin{figure}[ht]
    \centering
    \includegraphics[width=\linewidth]{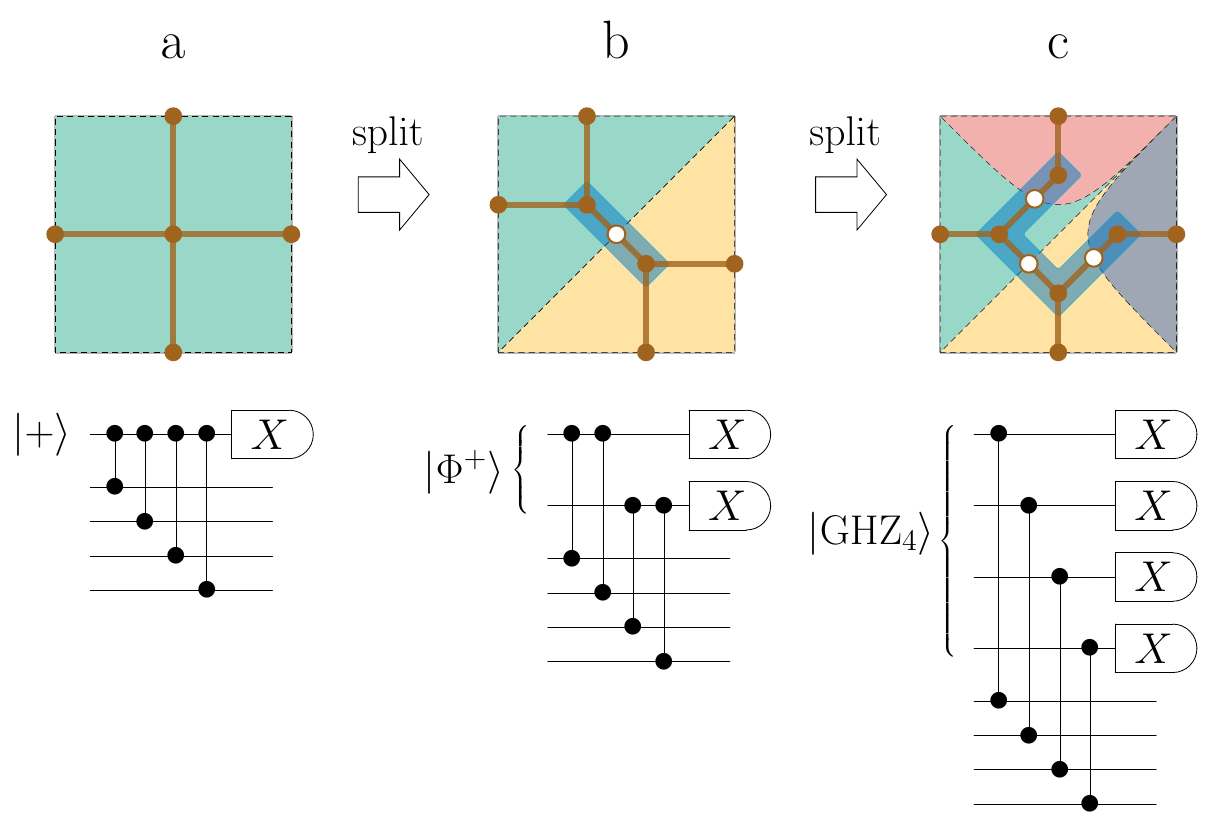}
    \caption[Face splitting on a square.]{Face splitting on a square.
    \textbf{(a)}~Without a split, the square represents a monolithic cluster state without distributed entanglement. The corresponding circuit of $\plusket$ state initialization, C$Z$ gates, and Pauli-$X$ basis measurement is also drawn.
    \textbf{(b)}~A hybrid approach. The distributed face qubit is split over two parties represented by two faces, forming a bipartite GHZ or Bell state. The parity check is performed through two bilocal C$Z$ gates of each face to its neighboring qubits and two local measurements.
    \textbf{(c)}~A fully distributed approach. A face is split into four parties that share a GHZ state. Each party performs a single C$Z$ gate to its neighbor and performs a measurement.}
    \label{fig:face-splitting}
\end{figure}

Like a cell-vertex split, a face-edge split subdivides an existing face into two or more parts, adding new edges to separate the newly created faces. We give an example of this procedure on a square in Fig.~\ref{fig:face-splitting}. The newly created edges each support an additional qubit, always laying adjacent to two faces. Therefore, the subgraph supported by split edges and faces is a chain in the form of Fig.~\ref{fig:alternating-ghz}. We may replace the cluster state supported by the $n$ connected faces with an $n$-partite GHZ state. Because the removed qubits correspond to ``virtual'' $m=0$ measurement outcomes, their even parity plays no further role in the evaluation of error syndromes.

\begin{figure}[ht]
    \centering
    \includegraphics[width=\linewidth]{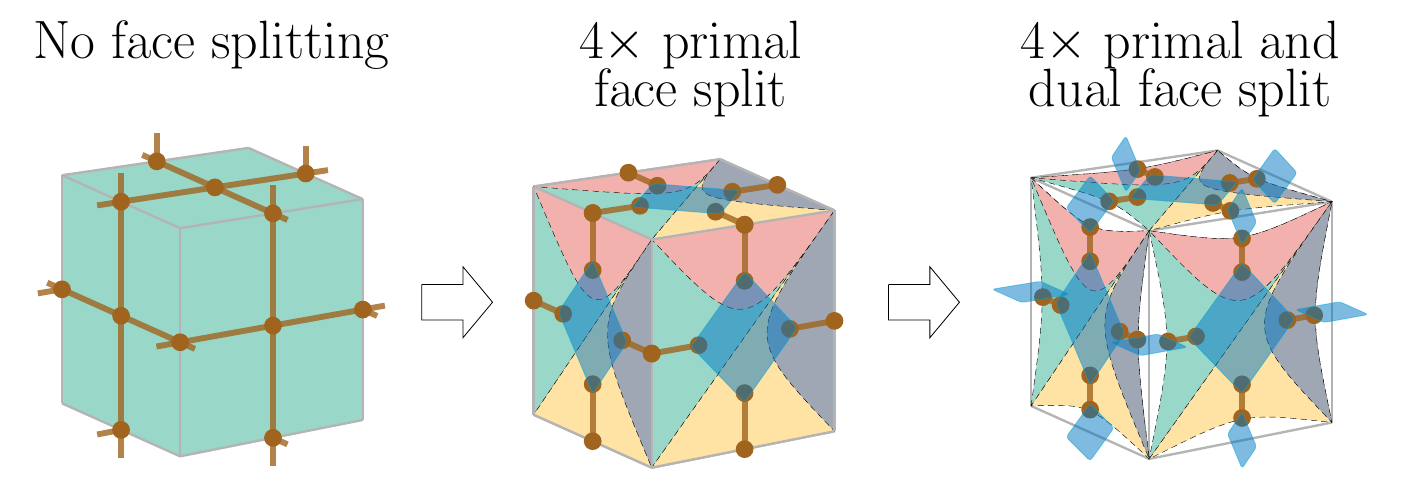}
    \caption[Face-edge splitting on a cubic cell.]{Face-edge splitting on a cubic cell. Starting from the left, a monolithic cluster state is transformed into a distributed cluster state by a four-partite split for each of its faces (see also Fig.~\ref{fig:face-splitting}). Performing the same procedure for dual faces (primal edges) produces a cluster state that is fully distributed, where each network node contains two cluster state qubits.}
    \label{fig:cubic-splitting}
\end{figure}

We may extend this procedure to both the primal and dual complex, producing GHZ states on both faces and edges. An example based on the cubic cluster state is given in Fig.~\ref{fig:cubic-splitting}. Starting from a monolithic architecture, a full 4-partite split of primal faces produces an architecture with nodes containing five qubits on a single edge and each of the adjacent split faces. Nodes are entangled with one another by GHZ states on every face. We may perform the same procedure for dual faces (primal edges), further reducing the number of qubits in each node to two.

We do not consider this in the rest of the paper but note that, instead of initializing the qubits introduced by a face-edge split as an entangled state and measuring them out individually, one can alternatively initialize these qubits regularly in $\ket{+}$ and measure them out with a joint \emph{(Type-II) fusion measurement}~\cite{browneResourceEfficientLinearOptical2005, griceArbitrarilyCompleteBellstate2011, ewert4efficientBellMeasurement2014}. This provides one with a method to transform (fault-tolerant) cluster states into so-called \emph{fusion networks} that form the basis of \emph{fusion-based quantum computing}~\cite{bartolucciFusionbasedQuantumComputation2023}.

\subsection{Circuit-level and network noise}\label{sec:entanglement}
In this section, we describe the noise models used for monolithic and distributed threshold calculations. For the monolithic simulations, the entire cluster state is built from $\plusket$ state preparation, followed by C$Z$ gates between every connected face-edge pair in the cluster state, concluded with a Pauli-$X$ basis measurement of every qubit. In this model, we do not consider the effects of memory decoherence. For circuit-level noise, the following noise sources are included:
\begin{enumerate}
    \item Noisy state preparation as a classical mixture $(1-p_\textrm{p})\ketbra{+} + p_\textrm{p}\ketbra{-}$. That is, with probability $p_\textrm{p}$, apply a phase-flip error on $\plusket$.
    \item Two-qubit depolarizing noise following every C$Z$ gate. That is, apply the channel $D(\rho)=(1-p_\textrm{g})\rho + p_\textrm{g}/15\sum_{A,B}(A \otimes B)\rho(A \otimes B)^\dagger$, where $(A,B) \in \{\mathbb{I}, X, Y, Z\}^2\setminus(\mathbb{I},\mathbb{I})$ is a pair of Pauli matrices excluding identity.
    \item Classical bit-flips following every Pauli-$X$ basis measurement. That is, a measurement outcome $m$ is flipped to $1-m$ with probability $p_\textrm{m}$. The corresponding noise channel takes on the form $\widetilde{P}^\pm(\rho) = (1-p_\textrm{m})P^\pm \rho\, P^\pm + p_\textrm{m} P^\mp \rho\, P^\mp$, where $P^\pm \equiv \ketbra{\pm}$ are the desired projectors in the Pauli-$X$ basis.
\end{enumerate}

For the entangled states used in the distributed simulations, we assume that parties have shared access to Bell states in the \emph{Werner} or \emph{isotropic} form
\begin{equation}\label{eq:werner_state}
\begin{split}
    \rho_{p_\textrm{n}} = & \left(1 - p_\textrm{n}\right)\ketbra{\Phi^+}{\Phi^+} + \frac{p_\textrm{n}}{3}\ketbra{\Phi^-}{\Phi^-} \\
             &+ \frac{p_\textrm{n}}{3}\ketbra{\Psi^+}{\Psi^+} + \frac{p_\textrm{n}}{3}\ketbra{\Psi^-}{\Psi^-},
\end{split}
\end{equation}
where $\ket{\Phi^+}=(\ket{00}+\ket{11})/\sqrt{2}$, $\ket{\Phi^-}=Z\ket{\Phi^+}$, $\ket{\Psi^+}=X\ket{\Phi^+}$, and $\ket{\Psi^-}=X \otimes Z\ket{\Phi^+}$. We refer to $p_\textrm{n}$ as the parameter that describes ``network noise''. In this paper, we do not take into account specific physical systems. We justify using Werner states by noting that a depolarizing channel (\textit{i.e.}, the most general noise channel) converts a perfect Bell state to a Werner state. Additionally, Werner states can be considered a general proxy for non-perfect Bell states, because every Bell state can be \emph{twirled}~\cite{Bennett1996b} into a Werner state using local operations and classical communication.

\begin{figure}[ht]
    \centering
    \includegraphics[width=0.75\linewidth]{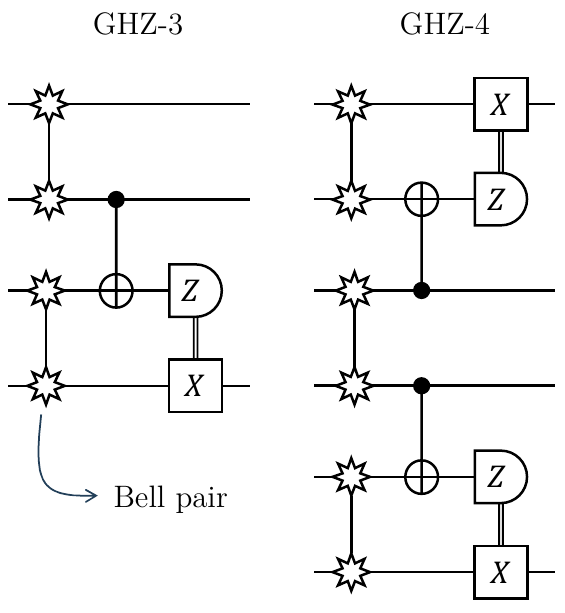}
    \caption[Fusion circuits for 3- and 4-partite GHZ states.]{Fusion circuits for 3- and 4-partite GHZ states. In each of the circuits, Bell states are fused through the application of local C$X$ gates and a subsequent $Z$ basis measurement of the target bits. The conditional bit-flips ensure that GHZ states have the desired form $\ket{0\dots0} + \ket{1\dots1}$.}
    \label{fig:ghz_creation}
\end{figure}

To generate GHZ states from these Bell states, we use the very straightforward method based on local parity measurements~\cite{Nickerson2015}. Fundamentally, a GHZ state may be created from Bell pairs by a local projective measurement of the $ZZ$ parity between two halves of two pairs shared by multiple parties. The circuits to create a $3$-partite GHZ from 2 Bell states and a $4$-partite GHZ state from 3 Bell states are drawn schematically in Fig.~\ref{fig:ghz_creation}.

We note that the circuits in Fig.~\ref{fig:ghz_creation} do not include distillation. Better quality GHZ states can be generated if the Bell states used to carry out the projective measurement are pre-distilled, or if GHZ states are distilled after creation. This, however, introduces a probabilistic factor to the GHZ creation protocols, which leads to higher numerical complexity in simulating the circuits.

\subsection{Numerical tools and considerations}\label{subsec:numerical_tools_and_considerations}
To decode error syndrome graphs, we have implemented a version of the \emph{Union-Find} (UF) decoder~\citep{delfosseAlmostlinearTimeDecoding2021}. This decoder is particularly attractive given its almost-linear time complexity and ease of implementation for both Pauli and erasure errors. Despite being a sub-optimal decoder, the phenomenological threshold for the cubic cluster state ($2.6\%$) is very close to that of the \emph{Minimum-Weight Perfect-Matching} (MWPM) decoder ($2.9\%$)~\citep{nickersonMeasurementBasedFault2018}, which is in turn not far from the optimal threshold ($3.3\%$)~\citep{ohnoPhaseStructureRandomplaquette2004}. Furthermore, since the Union-Find decoder is maximum-likelihood over the erasure channel (due to its built-in \emph{peeling} decoder~\citep{delfosseLineartimeMaximumLikelihood2020a}), we expect this performance gap to shrink further over noisy channels that are a combination of Pauli and erasure errors. We have specifically implemented the ``weighted-growth'' version of the Union-Find decoder, as introduced in the original Union-Find manuscript~\citep{delfosseAlmostlinearTimeDecoding2021}. Our implementation can be found in the repository of Ref.~\cite{vanmontfortDataSoftwareUnderlying2022}.

Error models per unit cell are constructed with a circuit simulator. Unit cells are defined according to the description in Sec.~\ref{subsec:unit_cell_and_crystal}, together with the splitting methods of Secs.~\ref{subsec:splitting-1} and~\ref{sec:splitting-2}. The simulator is implemented as a classical efficient stabilizer simulator. 
Per operation, the simulator applies a full error channel as a series of Pauli operators, ending with a measurement in the $n$-qubit Pauli basis. The choice for Pauli noise is justified in these simulations because the input states are (convex combinations of) stabilizer states, the operation noise is typically Pauli noise and measurement noise is described by classical bit-flips. In App.~\ref{subsec:characterization_of_noisy_channels}, we describe the details of constructing an error channel. Pauli twirling a state at the end of a series of non-Clifford operations is equivalent to Pauli twirling the individual operations before applying them. This allows us to also use the simulator in situations where one is interested in the Pauli-twirled version of the full error channel---in that case, it suffices to twirl the individual channel components before applying them.

The calculated error models are used to sample errors in Monte Carlo style on qubits of the full crystalline cluster state. These cluster states are constructed from the associated unit cell complex according to the crystal embedding procedure described in App.~\ref{subsec:crystal_embeddings}. Per Monte Carlo sample, we decode the syndrome graph and assign a logical failure whenever there is a logical error for a single pair of logical $X$ and $Z$ operators of the channel---see Sec.~\ref{subsec:cluster_error_correction} for more details.

\section{Results}\label{sec:results}
\begin{figure*}
    \centering
    \includegraphics[width=0.9\textwidth]{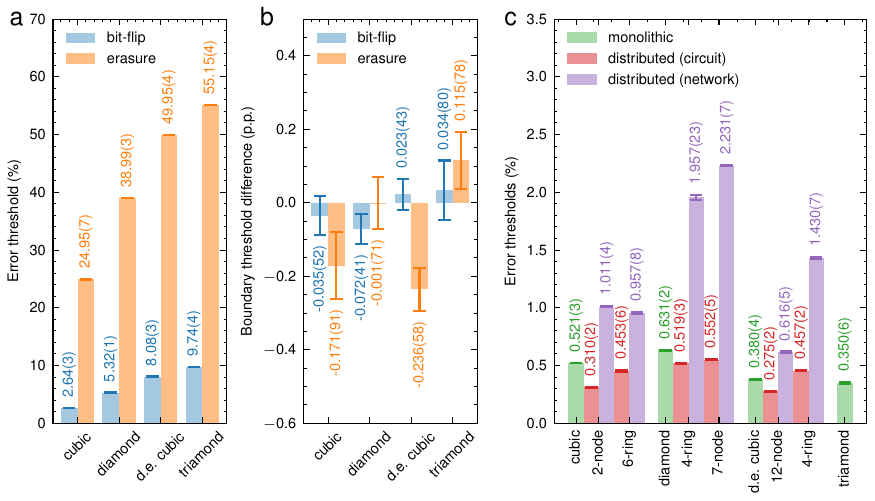}
    \caption{Overview of the threshold values found in our study, for different lattices and noise models considered.
    \textbf{(a)}~Summary of numerical thresholds for bit-flip and erasure-type phenomenological noise models of the four lattices described in this work.
    \textbf{(b)}~Differences of estimated threshold values for the same lattices and error models as in (a) with boundaries, expressed in percentage points (p.p.).
    \textbf{(c)}~Monolithic thresholds are gauged against circuit-level errors with a common error rate as defined in Secs.~\ref{sec:entanglement} and \ref{subsec:monolithic_thresholds}, and similarly for distributed thresholds gauged under only circuit-level noise. Network thresholds are obtained from the model as defined in Sec.~\ref{sec:entanglement}. For the monolithic thresholds, we show only the highest threshold value found with alternative gate orderings (see Sec.~\ref{subsec:monolithic_thresholds} in the main text for details).
    Numbers above the bars indicate the threshold, with $95\%$ confidence intervals around the value indicated as error bars and as decimals in parentheses.}
    \label{fig:threshold_results}
\end{figure*}

The results are organized into three parts, where the noise models become increasingly complex. A summary of the most important thresholds found can be found in Fig.~\ref{fig:threshold_results}. 

Secs.~\ref{subsec:phenomenological_thresholds} and \ref{subsec:phenomenological_thresholds_with_boundaries} provide thresholds for various geometries under a phenomenological noise model. In these sections, we reproduce earlier-known results and investigate the influence of a lattice boundary. For the phenomenological noise model, cluster states based on lattices that have a lower vertex degree are more resilient against errors. However, these cluster states typically require more two-qubit gates to construct. This aspect is not regarded by the phenomenological noise model. 

To investigate the trade-off between noise resilience and noise introduced by constructing the cluster state, we investigate scenarios with circuit-level and network noise in Secs.~\ref{subsec:monolithic_thresholds} and~\ref{subsec:distributed_thresholds}.  Sec.~\ref{subsec:monolithic_thresholds} discusses numerical thresholds for monolithic (\textit{i.e.}, non-distributed) architectures. These results are compared with the circuit-based error models for cluster states on a distributed network in Sec.~\ref{subsec:distributed_thresholds}. 

In the last part of this section, in Sec.~\ref{subsec:trade_off_GHZ_erasure}, we investigate how the GHZ success probability of distributed networks can be traded off against a higher erasure probability to achieve fault-tolerance against larger infidelity of the entangled states used. 

\subsection{Phenomenological thresholds}\label{subsec:phenomenological_thresholds}
\begin{figure*}
    \centering
    \includegraphics[width=0.85\textwidth]{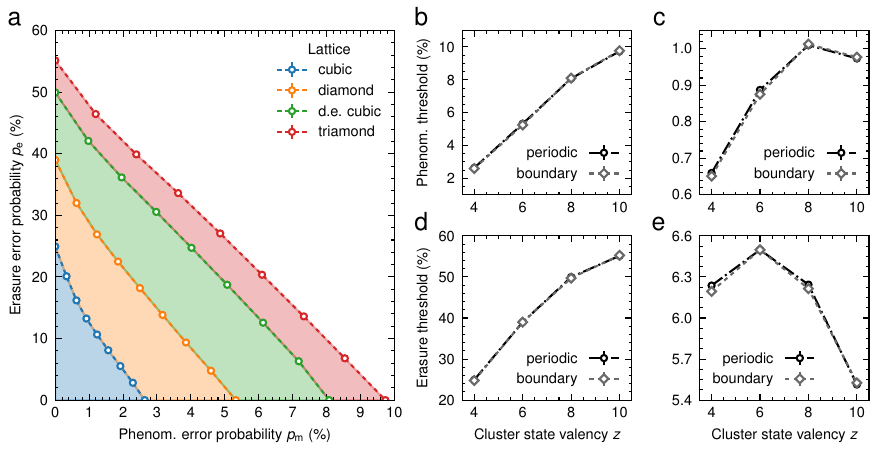}
    \caption[Fault-tolerant regions of various lattices under phenomenological bit-flip and erasure noise.]{Fault-tolerant regions of the cubic, diamond, double-edge (d.e.) cubic, and triamond lattices under phenomenological bit-flip and erasure noise.
    \textbf{(a)}~Regions are estimated based on individual threshold values.
    \textbf{(b)}~Thresholds $p_\textrm{m,th}$ under pure bit-flip noise as a function of cluster state valency. From left to right is the cubic, diamond, double-edge cubic, and triamond lattice. We provide thresholds with and without lattice boundaries---see Sec.~\ref{subsec:phenomenological_thresholds_with_boundaries} for more details. 
    \textbf{(c)}~Thresholds $p_\textrm{m,th}/z$ under pure weighted bit-flip noise, where $z$ is the cluster state valency.
    \textbf{(d)}~Thresholds $p_\textrm{e,th}$ under pure erasure noise. Results agree with known percolation thresholds for the given lattices.
    \textbf{(e)}~Thresholds $p_\textrm{e,th}/z$ under pure weighted erasure noise. 
    }
    \label{fig:pe-ft}
\end{figure*}

Before considering more realistic noise models, we first numerically evaluate erasure and phenomenological thresholds for several known lattices, which were constructed using cell-vertex splitting. These are the cubic, diamond, triamond, and so-called double-edge cubic cluster states. We go beyond the results in Ref.~\cite{nickersonMeasurementBasedFault2018} by determining \emph{fault-tolerant regions} against erasure and phenomenological errors. The results are depicted in Fig.~\ref{fig:pe-ft}. We consider a phenomenological error model. It corresponds to perfect state preparation of the cluster state followed by measurements that fail to report (\textit{i.e.}, erase) an outcome with probability $p_\textrm{e}$ or flip the outcome with probability $p_\textrm{m}$, both of which are i.i.d. 
In Fig.~\ref{fig:pe-ft}, we estimate a fault-tolerant region for the erasure probability and bit-flip probability. 
Data points of the fault-tolerant regions are calculated by sweeping over both error probabilities while keeping their ratio fixed. We use a different constant of proportionality at each data point on the fault-tolerant boundary. 

The isolated thresholds found for both types of noise can be found in Fig.~\ref{fig:threshold_results}a. 
The thresholds reported are slightly higher than those reported in Ref.~\cite{nickersonMeasurementBasedFault2018}. This can be attributed to a slightly different implementation of the Union-Find decoder---see Sec.~\ref{subsec:numerical_tools_and_considerations} for more details. 
Because phenomenological noise does not take into account higher error rates that arise with increasingly complex preparations of a cluster state, a fairer comparison is made by weighing each qubit error probability with the cluster state valency $z$, producing i.i.d.~noise with probabilities $zp_\textrm{e}$ and $zp_\textrm{g}$ instead. Such a model is called \emph{weighted phenomenological} error model in Ref.~\cite{nickersonMeasurementBasedFault2018}. Because the cubic, diamond, triamond, and double-edge cubic lattices are regular with valency $z\in\{4, 6, 8, 10\}$ on all qubits respectively, weighted thresholds may be calculated from unweighted thresholds by simply dividing by $z$.

\subsection{Phenomenological thresholds with boundaries}\label{subsec:phenomenological_thresholds_with_boundaries}
The lattices considered above repeat periodically in all three spatial directions. It may be difficult to prepare such a cluster state if we take into account the connectivity of the qubits. We gauged how the performance of a cluster state is affected by the introduction of boundaries, under the same phenomenological noise model of both bit-flips with probability $p_\textrm{g}$ and erasure errors with probability $p_\textrm{e}$. Each lattice is introduced to a smooth boundary along the $x=0$ plane, and a rough boundary along the $y=0$ plane. For the smooth boundary, we remove elements of the correlation surface defined by a \emph{dual} logical membrane and its closure and introduce a boundary on the remaining dangling edges in the dual complex. The rough boundary is introduced in the same way, by swapping primal and dual notions. 

\begin{figure*}
    \centering
    \includegraphics[width=0.85\textwidth]{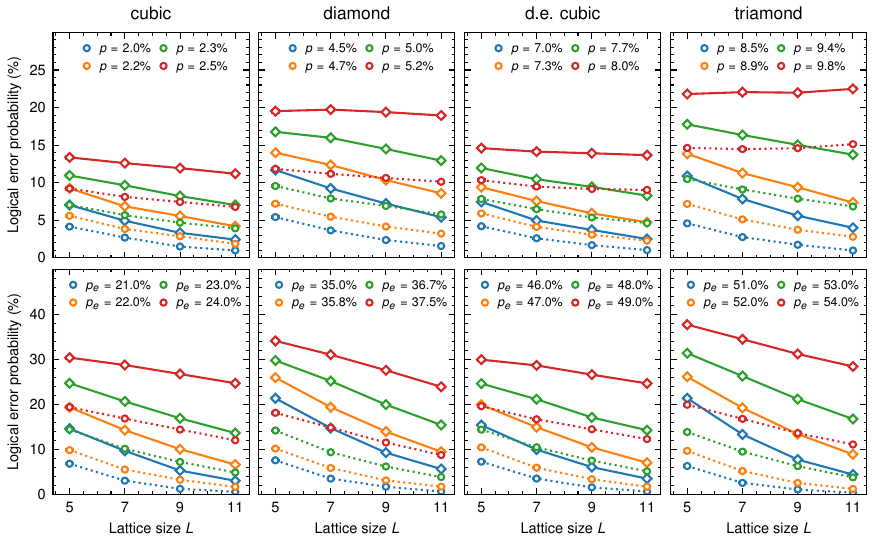}
    \caption[Sub-threshold logical error rate scaling of various unit cells.]{Sub-threshold logical error rate scaling of cubic, diamond, double-edge (d.e.) cubic, and triamond lattices with and without boundaries for phenomenological bit-flip (top row) and erasure (bottom row) noise models. Data for periodic conditions has round markers with a dashed line. Although the thresholds are minimally affected, sub-threshold error probabilities of any lattice with a rough and smooth boundary are about $1.5-2$ times the rate for the same lattice under periodic boundary conditions.}
    \label{fig:subthreshold-boundaries}
\end{figure*}

In Figs.~\ref{fig:threshold_results}b and~\ref{fig:pe-ft}b-e, we show how the threshold values for the phenomenological and erasure noise models change with the introduction of the boundaries. The results show that differences in the phenomenological bit-flip threshold values are insignificant. For the erasure thresholds, only the cubic and double-edge cubic lattices are slightly affected. The results indicate that the introduction of boundaries affects the thresholds only minimally, at least for the lattices and noise models considered here. However, we found that boundaries \emph{do} impose significantly weaker \emph{sub-threshold scaling}. The sub-threshold scaling is the rate at which the logical error rate is suppressed below the threshold. Fig.~\ref{fig:subthreshold-boundaries} shows that boundaryless architectures have a favorable error probability suppression that is about $50-100\%$ as effective as the same architecture with boundaries. These results are consistent with recent work, where it was also shown that the introduction of boundaries leaves the threshold nearly invariant, but negatively impacts error rate scaling by roughly the same factor~\cite{bombinLogicalBlocksFaultTolerant2023}. 

\subsection{Monolithic thresholds}\label{subsec:monolithic_thresholds}
To compare distributed thresholds against monolithic implementations of the same geometry, we first benchmark monolithic architectures with the circuit-level noise models from Sec.~\ref{sec:entanglement}. In the results that follow, we set all probabilities $p_\textrm{p}=p_\textrm{g}=p_\textrm{m}\equiv p_\textrm{o}$ equal (see Sec.~\ref{sec:entanglement} for the definition of the parameters), and sweep a threshold over the value of $p_\textrm{o}$. What remains is a specification of the ordering of C$Z$ gates in the circuits, since pure C$Z$ gates all commute with one another, whereas their noisy versions do not. Because no qubit can interact with two C$Z$ gates simultaneously, a valid ordering may be extracted from an edge coloring of the corresponding $\bound{2}$ boundary map of the cluster state. This graph is bipartite by definition. Therefore, the chromatic index (\textit{i.e.}, the minimum number of colors needed for an edge-coloring) equals the maximum degree of any vertex in the graph---\textit{i.e.}, the maximum valency of the cluster state~\cite{konigGraphokEsMatrixok1931}. The cubic, diamond, double-edge cubic, and triamond lattices are all regular, and so their chromatic indices are 4, 6, 8, and 10, respectively. 

For a chromatic index $i$ and a given coloring, there are $i!$ different ways to order the edges and thus the C$Z$ gates. Rotational and reflection symmetries of the cubic cluster state imply that there are only two colorings that correspond to a unique sequence of gates: a ``(counter)clockwise'' sequence going around a face, and a ``zigzag'' sequence jumping to opposite sides first. For diamond, double-edge-cubic and triamond lattices, the number of orderings quickly explode as $6!=720$, $8!\approx4\times 10^4$ and $10!\approx3.6\times 10^6$ (not taking into account symmetries). We have not included an exhaustive search of all orderings and their corresponding thresholds up to symmetries, but only investigate a subset of orderings.

\begin{figure*}
    \centering
    \includegraphics[width=0.9\textwidth]{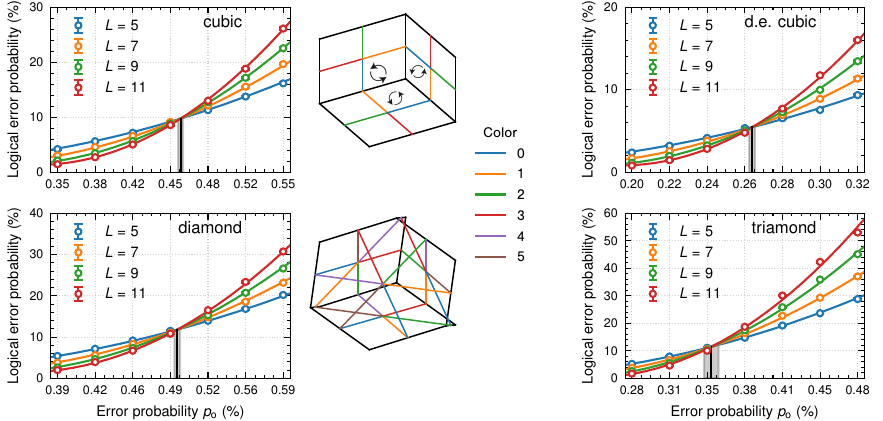}
    \caption[Monolithic thresholds.]{Monolithic thresholds $p_\textrm{o,th}$ for specific gate orderings. The thresholds are calculated with, for the cubic and diamond unit cell, a straightforward order of C$Z$ gates according to a ``(counter)clockwise'' coloring of the diagram on the right (see arrows in the diagram for the cubic lattice). We omit details on (counter)clockwise orderings for the double-edge cubic and triamond lattice. Gates of a single color are performed simultaneously for all unit cells in the lattice. Each data point constitutes $50 000$ samples. Error bars of the logical error rate are given as $95\%$ confidence intervals but are too small to be discernible in most cases. The threshold value is highlighted with a $95\%$ confidence interval based on its least-squares estimate of a second-order polynomial of the logical error rate around the threshold value---see App.~\ref{app:statistical_error} for details.}
    \label{fig:monolithic-thresholds}
\end{figure*}

Monolithic thresholds with the (counter)clockwise orderings are shown in Fig.~\ref{fig:monolithic-thresholds}. An overview of all monolithic thresholds is included in Fig.~\ref{fig:threshold_results}c. For the cubic lattice, a zigzag ordering of C$Z$ gates slightly outperforms the (counter)clockwise ordering. The diamond lattice outperforms a cubic cluster state, whereas the double-edge cubic lattice drops in performance for the orderings we considered. In all cases, (counter)clockwise orderings of the gates (at least in the primal complex) tend to produce lower thresholds than orderings that skip multiple edges at a time. We can intuitively understand this result by considering that a single Pauli-$X$ error on a face qubit spreads through all subsequent C$Z$ gates as correlated Pauli-$Z$ errors to neighboring edge qubits. In a (counter)clockwise ordering, the Pauli-$Z$ errors produce a single strand that wraps around the face. 
In a zigzag orientation, $Z$ error strings can form disconnected chains, such that a single Pauli-$X$ error produces multiple pairs of syndromes on different sides of the face. 
Initial numerical analysis shows that (counter)clockwise orderings indeed seem to result in, on average, longer error strings. 
The triamond lattice was gauged for a single ordering due to its complexity and performed poorly.

In the limiting case that all gate errors are Pauli-$Z$ errors, the relative impact of gate order disappears, and thresholds coincide with the weighted phenomenological thresholds considered above. In Ref.~\cite{newmanGeneratingFaultTolerantCluster2020}, Newman \textit{et al.} consider an error model that, after each C$Z$ gate, applies an $X$-type error on a (primal) face qubit with probability $p_X$ and a $Z$-type error on a (primal) edge qubit with probability $p_Z$. Their results show that the best-performing lattice in terms of threshold moves from higher to lower valency as Pauli-$X$ errors start to dominate. Our results show a similar tendency under depolarizing gate noise for the higher valent double-edge cubic lattice and triamond lattices when compared to the lower valent cubic and diamond lattices.
We can summarize this finding by stating that as the cluster state valency increases, depolarizing noise incurs a larger cost on the threshold value. Under these noise models, there exists an optimal threshold resulting from a trade-off between the complexity of the geometry of the cluster state, and the structure of the noise created by the circuit. Noise bias is one example where this trade-off may be abused, by taking advantage of the architecture of the cluster state in either primal or dual lattice structures.

The monolithic cluster state thresholds do not compete with surface code monolithic thresholds, which are estimated at $0.90\%$ and $0.95\%$ under the same noise model~\cite{Nickerson2013}. It should be noted that the numbers for all non-cubic lattices provided here are likely sub-optimal, as we have only gauged the performance for a subset of C$Z$ gate orderings. Nevertheless, estimates show that cluster states defined on the diamond lattice can outperform the cubic cluster state in the presence of circuit-level noise. For cluster states with even higher valencies, the cost of initialization negatively impacts the value of the threshold, consistent with the results under weighted phenomenological noise models. These results warrant further optimizations of the gate orderings and comparisons with other lattices.

\subsection{Distributed thresholds}\label{subsec:distributed_thresholds}
\begin{figure*}
    \centering
    \includegraphics[width=0.93\textwidth]{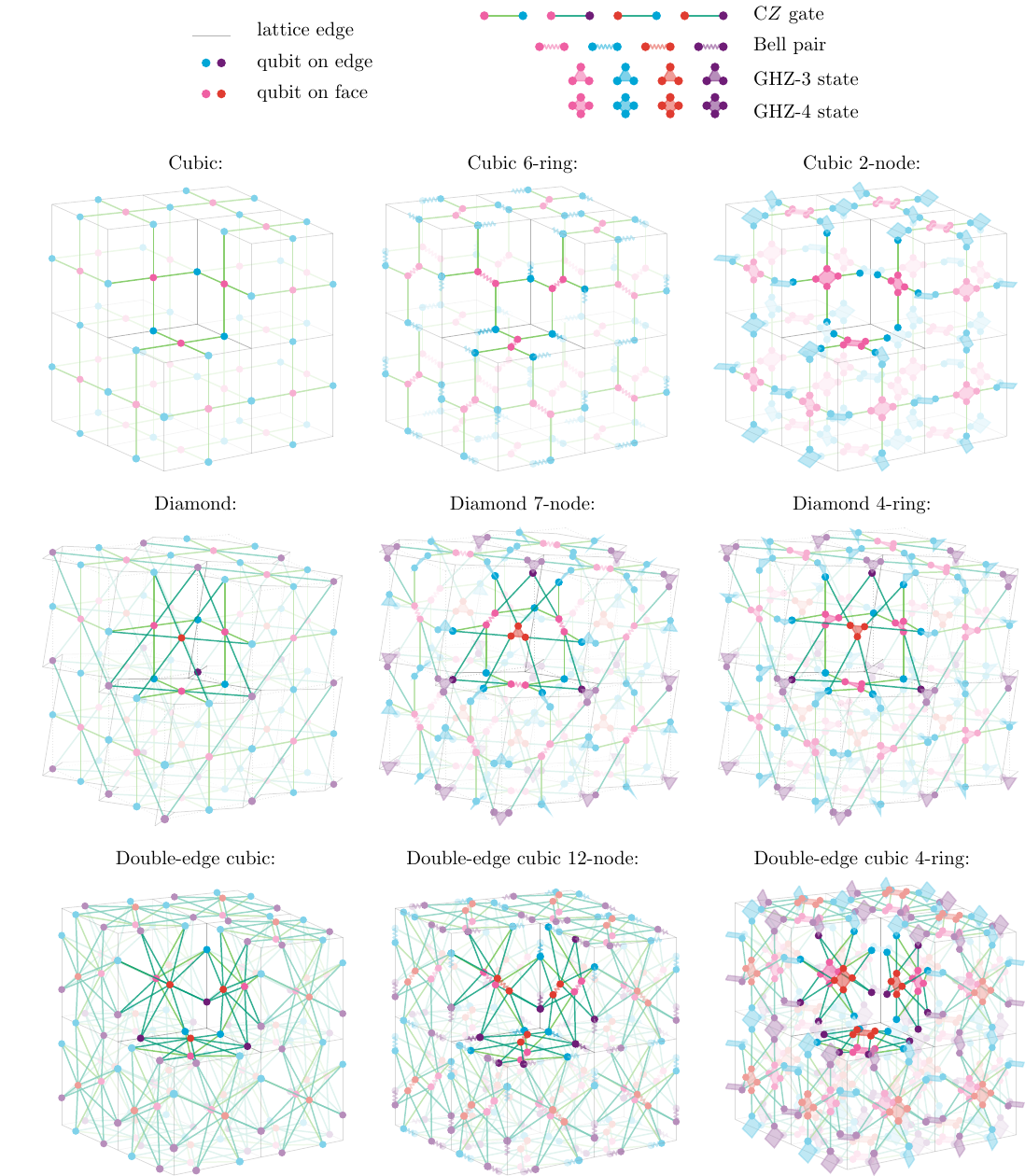}
    \caption[Distributed architectures for various lattices.]{Distributed architectures for various lattices, obtained through splitting the faces of the monolithic cluster state. A single-face split leads to a Bell state, whereas an $n$-split produces a GHZ state. They form connected components that are distinct nodes in a distributed network. We identify six different architectures: two each for the cubic, diamond, and double-edge cubic lattices. The left column shows purely monolithic cluster states. In the middle column, cluster states are initialized with Bell states on every face and edge, except for the diamond lattice that also contains weight-3 GHZ states. Architectures in the right column are initialized with 4-partite GHZ states for the cubic and double-edge cubic lattices, and 3-partite states for the diamond lattice. Architectures in the middle column tend to have larger nodes than those in the right column.}
    \label{fig:distributed-architectures}
\end{figure*}

In this section, we investigate thresholds for distributed implementations of the cluster states. We use the face-edge splitting operation of Sec.~\ref{sec:splitting-2} to split faces of the cubic, diamond, and double-edge cubic lattices. The cubic lattice was gauged for two different splits: one along the diagonals, producing network nodes with six cluster qubits in a ring and entangled through Bell states, and one on the entire face, producing 2-qubit nodes that are entangled through 4-partite GHZ states. Both these architectures also appear in the context of fusion-based quantum computation in Bartolucci \textit{et al.}~\cite{bartolucciFusionbasedQuantumComputation2023}. The structure of the diamond lattice is more intricate, and we produce two different architectures that contain only 3-partite GHZ states (the 4-ring) and an architecture with a mixture of Bell states and 3-partite GHZ states. Architectures for the double-edge cubic lattice resemble the cubic lattice: the first contains only Bell states, whilst the other shares 4-partite GHZ states. All architectures are drawn schematically in Fig.~\ref{fig:distributed-architectures}. We present an investigation on the stabilizer fidelity of the distributed unit cell implementations of Fig.~\ref{fig:distributed-architectures} in App.~\ref{sec:stabilizer_fidelities_for_distributed_setting}. 

In the distributed model, we set $p_\textrm{p}=p_\textrm{g}\equiv p_\textrm{o}$. On top of that, we use entangled states to connect the cluster states that are separated by the splits. For these states, we assume that nodes can prepare Werner states with fidelity $1-p_\textrm{n}$ and that GHZ states are generated with the protocols of Fig.~\ref{fig:ghz_creation}. 
In the absence of network links, this circuit-level model reduces to the model discussed in Sec.~\ref{subsec:monolithic_thresholds}. 

\begin{figure*}
    \centering
    \includegraphics[width=0.9\textwidth]{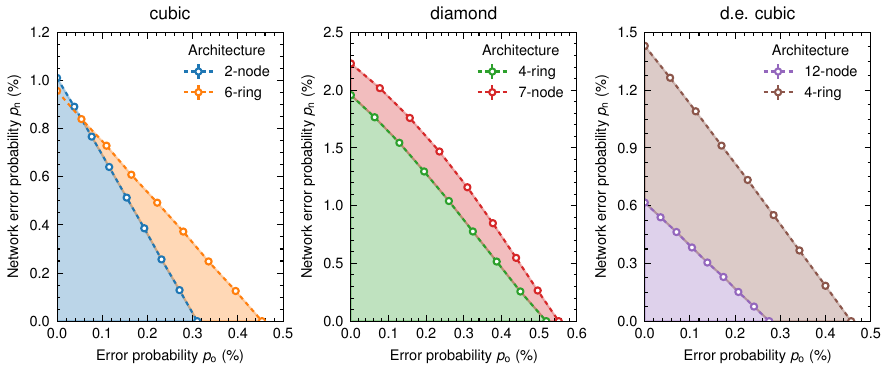}
    \caption[\textcolor{red}{Should there be $t$ somewhere?} Fusion-based thresholds of various lattices.]{Fusion-based thresholds for the six distributed architectures of the cubic, diamond, and double-edge cubic lattices. The circuit-level noise parameter $p_\mathrm{o}$ and network error rate $p_\textrm{n}$ are both swept, and a fault-tolerant region is estimated over thresholds of both parameters. See main text for simulation details.}
    \label{fig:fusion-thresholds}
\end{figure*}

Using this physical model, we numerically establish thresholds for the distributed lattices introduced above. In the same way as phenomenological bit-flip and erasure thresholds, we estimate fault-tolerant regions for error probabilities $p_\textrm{o}$ and $p_\textrm{n}$. 
The results are shown in Fig.~\ref{fig:fusion-thresholds} and in the overview of Fig.~\ref{fig:threshold_results}c.

These results show that network error rate thresholds for both the 2-node and 6-ring cubic designs are similar. Even though higher-valent GHZ states tend to produce lower-quality stabilizers, we suspect that the superior performance of the 2-node design may be due to the size of its node. Because the 2-node architecture has only a single C$Z$ gate per node, errors spread to only one adjacent qubit, as opposed to the 6-ring architecture that moves such an error to two adjacent qubits. Despite the more significant propagation of errors, the 6-ring architecture has a higher threshold against gate and measurement errors (the probability $p_\mathrm{o}$) than the 2-node design. Most likely, this is because measurement errors have a higher influence in the 2-node architecture, which has more qubits. 
Both diamond architectures outperform the cubic lattice by a factor of roughly two. This result is promising, especially considering that the depolarizing thresholds also outperform cubic architectures, as was already the case for the monolithic thresholds discussed above.

For the double-edge cubic architectures, network error rate thresholds drop again. Nevertheless, the 4-ring design with GHZ states outperforms the bigger 12-node design, which we may explain in the same way as in the cubic case: in the 4-ring lattice, errors propagate to fewer neighboring qubits compared to the 12-node architecture. An important difference with the cubic lattices is that the 4-ring design benefits enough from the lower error spreading to keep outperforming the 12-node design under pure gate and measurement noise, despite the additional qubit measurements in the 4-ring implementation.

Thresholds for the circuit-level noise rate $p_\textrm{o,th}$ of these architectures are not optimal, due to the way that the ordering of C$Z$ gates affects the threshold. 
We find that, for all architectures considered, the values found for $p_\mathrm{o,th}$ without network noise are similar to the monolithic thresholds. Similar to the earlier analysis of distributed designs, comparing distributed to monolithic architectures involves trading off less error propagation in the distributed architectures versus fewer measurement errors in the monolithic architecture.

\begin{figure*}
    \centering
    \includegraphics[width=0.73\textwidth]{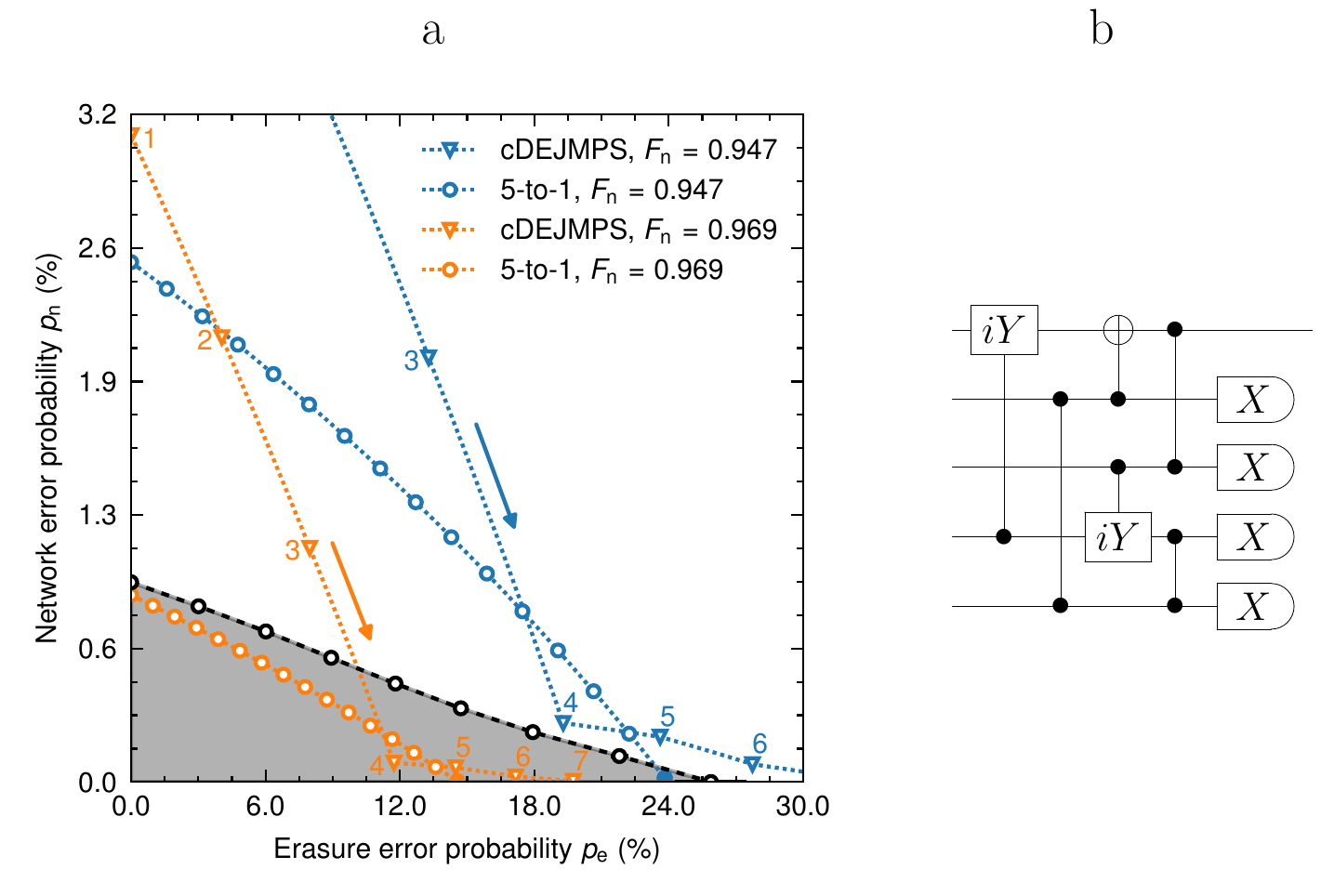}
    \caption[Trade-off between erasure and network error rate.]{\textbf{(a)}~Trade-off between erasure and network error rate. The fault-tolerant region of the six-ring architecture is drawn (gray), where entangling links have a network error rate $p_\textrm{n}$ and an independent probability $p_\textrm{e}$ to fail. Gates and measurements are assumed to be noiseless. A failed link causes the edge to be erased---in the absence of network error, we obtain the phenomenological erasure threshold of the cubic lattice. The colored lines reflect infidelities and failure rates of entanglement distillation protocols that use identical copies of a Werner state with initial fidelity $F_\textrm{n}$ to distill a single Bell state with higher fidelity---\textit{i.e.}, with lower network error rate. Distillation may bring a non-fault-tolerant architecture into the fault-tolerant regime. Data marked with triangles correspond to concatenated DEJMPS (``cDEJMPS'') distillation protocols~\cite{Deutsch1996}, where the numbers indicate how many Bell states are used to distill the final state. Data marked with colored circles is based on a distillation protocol (printed in panel b) that consumes five Bell states. For this protocol, the data point on the bottom right of each curve corresponds to a variant where, per Bell state, only coinciding measurement outcomes are accepted. The data point on the top left of each curve corresponds to a variant where one accepts all measurement outcomes. Intermediate protocols accept a subset of non-coinciding measurement results. \textbf{(b)}~Possible circuit for the 5-to-1 distillation protocol. The circuit is based on the decoding circuit of the 5-qubit error-correction code---see Ref.~\cite{goodenoughNeartermDistillationProtocols2023} for more details. Each line corresponds to half of a Bell state---\textit{i.e.}, the same circuit has to be applied to the qubits that form the other half of these Bell states.}
    \label{fig:sixring-erasure}
\end{figure*}

\subsection{Trade-off GHZ success rate and erasure probability}\label{subsec:trade_off_GHZ_erasure}
The results in the previous sections indicate that fault tolerance can be achieved with network error probabilities below $\sim2$\%---\textit{i.e.}, with Bell state fidelities above $\sim98$\%. This condition can be challenging from an experimental perspective. 
Fortunately, it is possible to boost the fidelity of entanglement before cluster state preparation through entanglement distillation. 
In this process, we can model distillation failures as erasures on the corresponding qubit(s), and use a suitable decoder, such as the Union-Find decoder, to deal with these qubit erasures. A suitable decoder can jointly correct errors and erasures. Phenomenological erasure thresholds can be roughly one order of magnitude higher than bit-flip error thresholds, so one may reasonably expect that trading in failed distillation attempts for higher-quality links will be worth the cost. 

We make this argument more quantitative in Fig.~\ref{fig:sixring-erasure}, using the cubic 6-ring architecture as an example. In this architecture, every Bell pair is successfully heralded with probability $1-p_\textrm{e}$ and network error rate $p_\textrm{n}$, and discarded with probability $p_\textrm{e}$. The fault-tolerant region is simulated in the same way as before, this time sweeping over the network and erasure error rates, where circuit-level noise is not taken into account---\textit{i.e.}, $p_\textrm{o}$ is set to $p_\textrm{o}=0$. We can apply an entanglement distillation protocol to distill each Bell state used in the cluster state by consuming multiple Bell states with initial fidelity $F_\textrm{n} = 1 - p_\textrm{n}$. Distillation protocols that we consider are (concatenated versions of) the DEJMPS protocol~\cite{Deutsch1996} and the 5-to-1 protocol~\cite{goodenoughNeartermDistillationProtocols2023} of Fig.~\ref{fig:sixring-erasure}b. The output fidelities and failure rates of these distillation protocols are shown in the graph for two different values of the initial fidelity $F_\textrm{n}$. For these protocols, there is a trade-off between fidelity and failure rate. Importantly, we can directly link this trade-off to the trade-off between erasure and network error rate---this makes it possible to move the state inside the fault-tolerant region. For the distillation protocols considered, the first crossing with this region happens for an initial network error probability of $p_\textrm{n} \approx 5\%$, which is a five-fold increase of the approximate 6-ring threshold value of $1\%$ without distillation. These results are particularly promising for the diamond architecture, where both the network error threshold of roughly $2\%$ and the erasure threshold of $40\%$ are higher than the cubic lattice.

\section{Conclusion}
In this paper, we have provided several tools and numerical analyses to explore fault-tolerant measurement-based quantum computing architectures built from smaller units, in the context of modular, distributed, or networked computing. This was achieved with a method that allows us to distribute a fault-tolerant cluster state over multiple parties. This method augments the splitting procedure of previous work by Nickerson and Bombín~\citep{nickersonMeasurementBasedFault2018}. In the distributed context, the resulting states will lead to entanglement in the form of a Bell or GHZ state, such that existing methods for state generation and distillation may be used to design such a fault-tolerant architecture. 

The performance of various three-dimensional cluster state architectures was studied through numerical evaluation of their fault-tolerant thresholds, which quantify the rate of specific sources of error below which fault-tolerant computation is possible. We find that the diamond lattice outperforms the traditional cubic cluster state for monolithic architectures suffering standard noise models---\textit{i.e.}, the noise models described in Sec.~\ref{sec:entanglement}. It should be mentioned that even better results may be achieved with different permutations of the entangling C$Z$ gates during cluster state preparation. In the cases we have considered, (counter)clockwise orderings of C$Z$ gates tend to produce lower thresholds. 

Furthermore, we have gauged the performance of the same lattices in a distributed setting. For designs based on the cubic lattice, error thresholds of the network noise are around $1\%$. We consider two different designs of distributed cluster states defined on top of a diamond lattice, which outperform cubic thresholds roughly by a factor of two. Using a cubic architecture, we show how entanglement distillation may bring a non-fault-tolerant design into the fault-tolerant regime by trading in network noise for erasure errors in the cluster state. Combined with favorable erasure thresholds of the diamond lattice, these results indicate that distributed fault-tolerant cluster states may outperform topological error-correction codes, and warrant additional numerical simulations of these distributed networks in the presence of entanglement distillation.

There are several potential avenues for further investigation. 
On the one hand, our circuit-based qubit error models are limited to depolarizing and erasure-type noise, but one may estimate fault-tolerant thresholds for models that more accurately represent errors in present-day quantum hardware. 
Furthermore, it would be interesting to consider protocols that cannot be described as a simple sequence of instructions but contain dependencies and branches that split based on intermediate decisions---as, \textit{e.g.}, protocols that contain entanglement distillation. Previous results in the field apply pre-calculated error models (\textit{i.e.}, quantum channels) on a unit cell level, and randomly sample from the error model during Monte Carlo threshold calculations~\cite{Nickerson2013, nickersonFreelyScalableQuantum2014}. This is possible because the fault-tolerant protocol consists of identical rounds that are applied over time, where each round is split up and grouped into multiple sub-rounds that act on disjoint subsets of qubits, so that the entire protocol may be simulated as distinct sections that are separated in both space and time. This is not necessarily the case for the three-dimensional cluster states that we consider here, in which case the entire cluster state should be simulated at a global level, i.e. without pre-calculating error models of individual sections~\cite{gidneyBenchmarkingPlanarHoneycomb2022}.

As an alternative to measurement-based quantum computation, one may consider so-called fusion-based quantum computation~\citep{bartolucciFusionbasedQuantumComputation2023}. It combines a low circuit depth with the topological features of cluster states. The fundamental operations in fusion-based quantum computation are resource-state generation and fusions. Fault-tolerant fusion-based architectures rely heavily on resilience against erasure, and the structures considered in this work are a natural candidate to consider in this alternative framework of computation.

\section*{Acknowledgments}
The authors would like to thank Michael Newman, Naomi Nickerson, Tim Taminiau, and Barbara Terhal for helpful feedback and/or discussions. We gratefully acknowledge support from the joint research program ``Modular quantum computers'' by Fujitsu Limited and Delft University of Technology, co-funded by the Netherlands Enterprise Agency under project number PPS2007. 
This work was supported by the Netherlands Organization for Scientific Research (NWO/OCW), as part of the Quantum Software Consortium Program under Project 024.003.037/3368. This work was partially supported by the JST Moonshot R\&D program under Grant JPMJMS226C. We thank SURF (www.surf.nl) for the support in using the National Supercomputer Snellius.

\appendix

\section{Quotient boundaries in unit cell complex}\label{subsec:unit_cell_complex}
By the prescription of Sec. \ref{subsec:unit_cell_and_crystal}, the unit cell complex is described as a sequence of vector spaces
\begin{center}
\begin{tikzcd}[row sep = tiny]
Q_3 \arrow{r}{\Qbound{3}{r}} & Q_2 \arrow{r}{\Qbound{2}{r}} & Q_1 \arrow{r}{\Qbound{1}{r}} & Q_0,
\end{tikzcd}
\end{center}
with each $Q_i$ over the field $\Ztwo$ and with quotient boundaries $\Qbound{i}{r}: Q_i \mapsto Q_{i-1}$. Similar to the equivalence between $\overline{C}_i$ and $C_{D-i}$, we define $\overline{Q}_i \cong Q_{D-i}$. One may verify that the dual quotient boundaries $\Qdualbound{i}{r}:\overline{Q}_i \mapsto \overline{Q}_{i-1}$ are related to primal boundaries as
\begin{equation}
\label{eq:dual-qbound}
    \Qdualbound{i}{r} = \left(\Qbound{D+1-i}{\overline{r}}\right)^T.
\end{equation}
Importantly, the translation vector $\mathbf{r}$ is also reversed to $\mathbf{\overline{r}}$. By dualizing boundary maps of unit cell complex directly, one may construct a representation of the dual unit cell without redefining it from the dual crystal. The zero map conditions $\bound{i-1}\bound{i} = 0$ take the form of
\begin{equation}
\label{eq:doubleqbound}
    \sum_{p}\Qbound{i-1}{p}\Qbound{i}{r-p}=0 \quad \forall\mathbf{r}.
\end{equation}
The proof is given in App.~\ref{subsec:crystal_embeddings}.

\begin{figure*}[t]
    \centering
    \includegraphics[width=0.7404\textwidth]{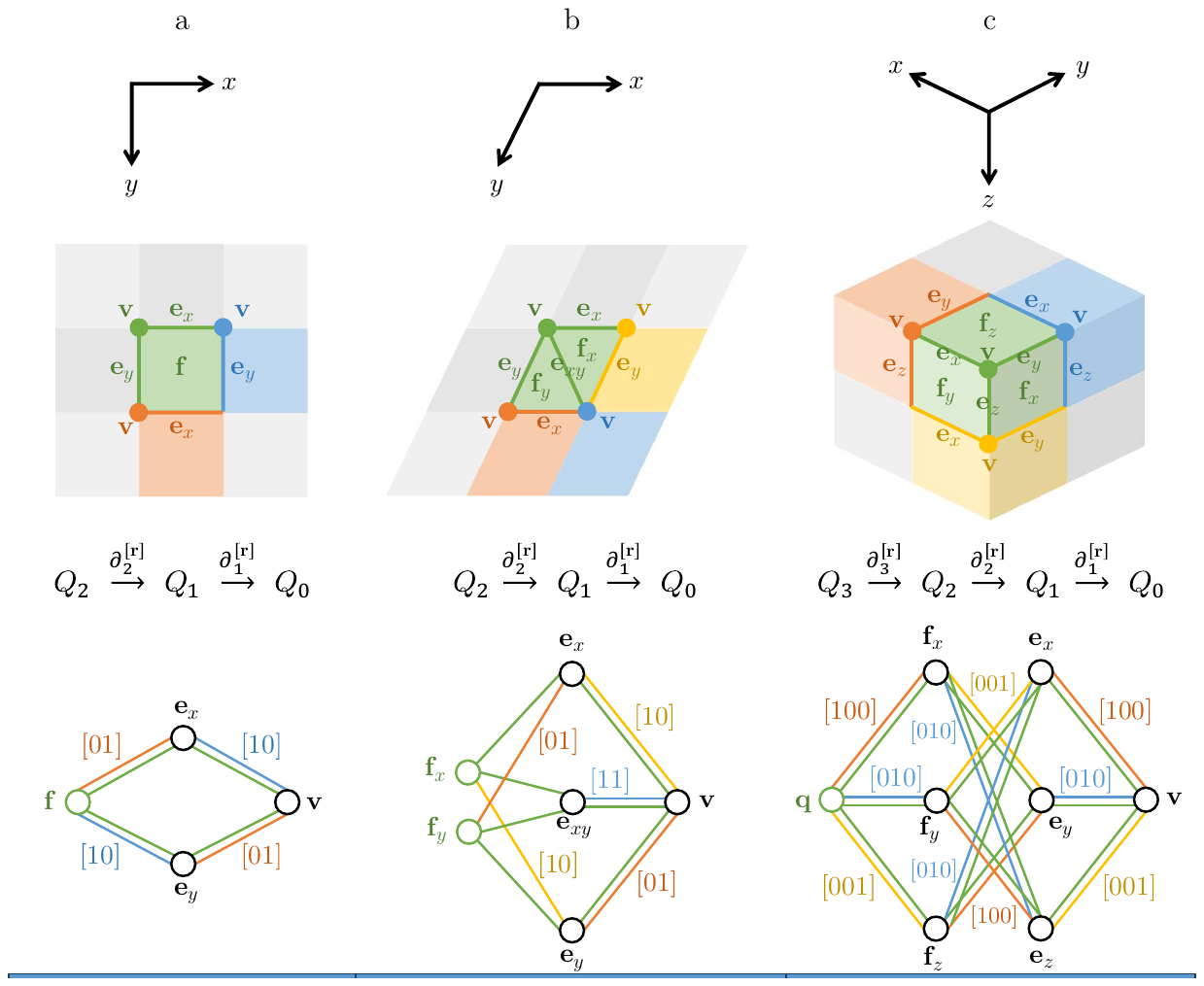}
    \caption[Unit cell complexes as a labeled graph.]{Unit cell complexes as a labeled graph. Unlabelled edges correspond to a Miller index containing only zeros.
    \textbf{(a)}~A square lattice. There is one face, two edges, and one vertex. The face $\mathbf{f}$ is connected twice to both $\mathbf{e}_x$ and $\mathbf{e}_y$ within the unit cell ($[00]$) and outside of it ($[01]$ and $[10]$, respectively). Because the complex is self-dual, similar relationships hold for its vertex $v$.
    \textbf{(b)}~A triangular lattice. This lattice can be created by splitting the faces of the square lattice: per unit cell, the triangular lattice has one extra face and one extra edge compared to the square lattice. From the asymmetry in the quotient boundary maps it is clear that this lattice is not self-dual.
    \textbf{(c)}~A cubic lattice. Per unit cell, there is one cell $\mathbf{q}$; three faces $\mathbf{f}_x$, $\mathbf{f}_y$, and $\mathbf{f}_z$; three edges $\mathbf{e}_x$, $\mathbf{e}_y$, and $\mathbf{e}_z$; and one vertex $\mathbf{v}$.}
    \label{fig:unit-cells}
\end{figure*}

It is convenient to represent the underlying unit cell complex as a \emph{labeled} graph, which is essentially a sparse representation of its boundaries as \emph{arcs} and basis elements as \emph{nodes}. (We use nomenclature \emph{nodes} and \emph{arcs} for such a graph, to make the distinction between vertices and edges of the chain complex.) Every basis element $\left(\chain[q]{i}\right)_n \in Q_i$ is mapped to a node $q_{i,n}$, with two nodes $q_{i,n} \rightarrow_{\Miller{r}} q_{i-1,m}$ connected by an $\Miller{r}$-labelled arc if the $mn$th matrix element of the quotient boundary $\Qbound{i}{r}$ equals one. The maps $\Qbound{i}{r}$ thus form the \emph{biadjacency} matrices between the nodes of $Q_i$ and $Q_{i-1}$. We note that this description, including its labeling, resembles the \emph{vector method} for describing three-periodic networks as a quotient graph~\cite{chungNomenclatureGenerationThreeperiodic1984}, except that the nodes of our quotient graph also represent higher-dimensional elements in a chain complex, such as edges, faces, and cells for a three-dimensional complex. Examples of the square, triangular, and cubic lattice are given in Fig.~\ref{fig:unit-cells}.

\section{Crystal embeddings}\label{subsec:crystal_embeddings}
\begin{figure}[ht]
    \centering
    \includegraphics[width=\linewidth]{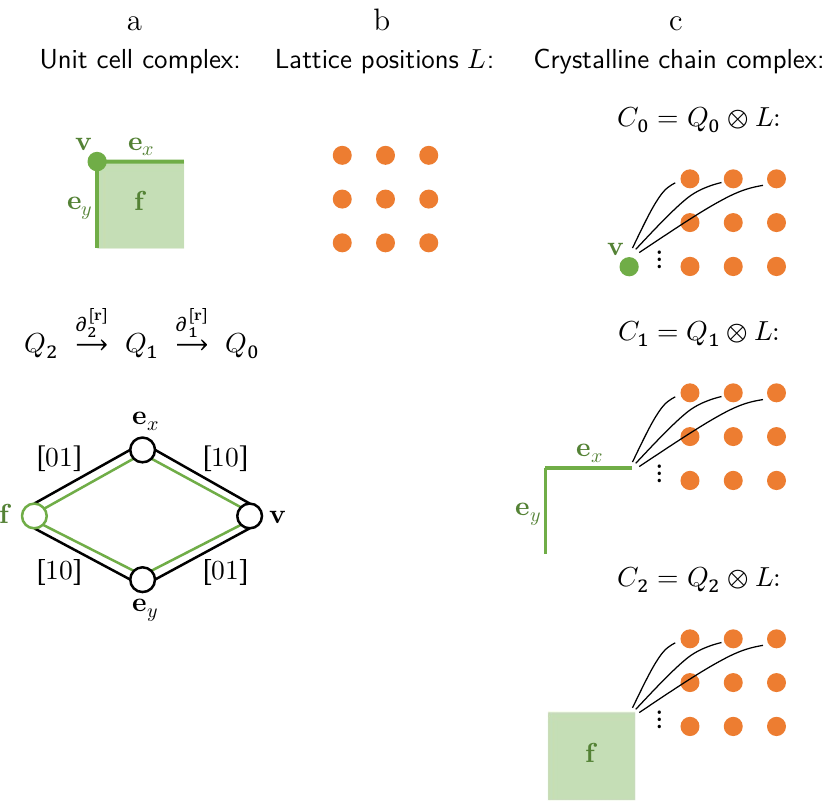}
    \caption[Graph product to create the full crystalline chain complex out of the unit cell complex.]{Abstract interpretation of how the full crystalline chain complex is created out of the unit cell complex introduced in App.~\ref{subsec:unit_cell_complex}. \textbf{(a)}~We use the example of the unit cell of the square lattice---see Fig.~\ref{fig:unit-cells} for more details. \textbf{(b)}~As described in Sec.~\ref{subsec:unit_cell_and_crystal} of the main text, we use an $N$-dimensional vector space $L\equiv\mathbb{Z}_2^{\oplus N}$ with as basis vectors the $N$ lattice positions of the lattice. \textbf{(c)}~The vector spaces $C_i$ of the full crystalline chain complex are realized with the \emph{graph product} between $L$ and the vector spaces $Q_i$ of the unit cell complex. The full boundary maps $\partial_i$ are constructed from the quotient boundary maps $\partial_i^{[\mathbf{r}]}$ according to Eq.~\eqref{eq:boundary_maps_out_of_quotient_boundary_maps}.}
    \label{fig:graph-product-unit-cell-complex}
\end{figure}

In Fig.~\ref{fig:graph-product-unit-cell-complex}, we show an intuitive interpretation of how the vector spaces $C_i$ of the full crystalline chain complex are constructed with the vector spaces $Q_i$ of the unit cell complex of App.~\ref{subsec:unit_cell_complex}. To construct the boundary maps $\partial_i$ of the full crystal from the quotient boundary maps $\partial_i^{[\mathbf{r}]}$, we let $\bound{i}^{\left(\mathbf{n}, \mathbf{m}\right)}: C_i \mapsto C_{i-1}$ be the boundary map that applies $\Qbound{i}{n-m}$ from cell $Q_i^{\left(m\right)}$ to cell $Q_{i-1}^{\left(n\right)}$ and is zero everywhere else:
\begin{equation}
    \bound{i}^{\left(\mathbf{n}, \mathbf{m}\right)} = \Qbound{i}{n-m} \otimes e_{nm}.
\end{equation}
Here, $e_{nm}: L \mapsto L$ is a \emph{matrix unit}, \textit{i.e.}, an $N \times N$ matrix with a one at indices $n, m$ and zero elsewhere. Then the boundary maps of the embedding are given as a sum
\begin{equation}
    \bound{i} = \sum_{m,n}\bound{i}^{\left(\mathbf{n}, \mathbf{m}\right)} = \sum_{m,n}\left(\Qbound{i}{n-m} \otimes e_{nm}\right).
\end{equation}
Because most $\Qbound{i}{n-m}$ are zero, we can substitute $\mathbf{r =n - m}$ and sum $\mathbf{r}$ only over the non-zero maps $\Qbound{i}{r}$, leading to
\begin{equation}
    \bound{i} = \sum_{m,r}\left(\Qbound{i}{r} \otimes e_{m+r,m}\right) = \sum_{r}\Qbound{i}{r} \otimes \left(\sum_{m}e_{m+r,m}\right).
    \label{eq:boundary_maps_out_of_quotient_boundary_maps}
\end{equation}
In the second equality, the sum over $\mathbf{m}$ may be carried over to the right by distributivity of the tensor product over addition. This last term is a permutation matrix with a single one in each row and column; it represents a \emph{translation} of a lattice point $\mathbf{m}$ to $\mathbf{m+r}$. Denote this term as $T_\mathbf{r}\equiv\sum_m e_{m+r,m}$, such that the embedding is given as
\begin{equation}
\label{eq:embedded-bound}
    \bound{i} = \sum_{\mathbf{r}}\Qbound{i}{r} \otimes T_\mathbf{r}.
\end{equation}
Intuitively, the crystal boundary is formed by ``gluing'' the boundaries between the unit cells $Q_i^{\left(\mathbf{m}\right)}$ and $Q_{i-1}^{\left(\mathbf{m+r}\right)}$ at lattice points $\mathbf{m}$ and $\mathbf{m+r}$ according to the map $\Qbound{i}{r}$, and repeating this process for every non-trivial quotient boundary map. The zero map conditions for quotient boundaries (Eq.~\eqref{eq:doubleqbound}) follow trivially. Because matrix units multiply as $e_{ij}e_{kl}=\delta_{jk}e_{il}$, the multiplication of two permutation matrices
\begin{equation}
\begin{split}
    T_\mathbf{p}T_\mathbf{q} & = \sum_{m,n}e_{m+p,m}e_{n+q,n} = \sum_{m,n} \delta_{m,n+q}e_{m+p,n} \\
	& = \sum_n e_{n+p+q,n} = T_\mathbf{p+q}
\end{split}
\end{equation}
represents the sum of their translations. Combining this result with the embedded boundaries (Eq.~\eqref{eq:embedded-bound}) directly, the composition of two maps equals
\begin{equation}
    \bound{i-1}\bound{i} = \sum_{p,q} \Qbound{i-1}{p}\Qbound{i}{q} \otimes T_\mathbf{p+q} = \sum_r \left(\sum_p \Qbound{i-1}{p}\Qbound{i}{r-p}\right) \otimes T_\mathbf{r},
\end{equation}
where we have substituted $\mathbf{r} = \mathbf{p} + \mathbf{q}$ in the last equality. This map is the zero map if and only if the term in brackets is zero for all $\mathbf{r}$, which is exactly the result stated in Eq.~\eqref{eq:doubleqbound}.

The above recipe for a crystal embedding may be expressed as a composition of \emph{direct products} between two graphs, given the following correspondences:

\begin{enumerate}
    \item The matrix $\Qbound{i}{r}$ is the biadjacency matrix of the subgraph $G[\Qbound{i}{r}]$ of the unit cell complex induced by edges with label $[\mathbf{r}]$. This definition is consistent with the graph description given in Sec.~\ref{subsec:unit_cell_complex}.
    \item The matrix $T_\mathbf{r}$ is the biadjacency matrix of the subgraph of the lattice $H[T_\mathbf{r}]$ induced by edges that translate each lattice point by $\mathbf{r}$. That is, each lattice point is represented by a node $\mathbf{m}$ and connected by an arc to the translated node $\mathbf{m} + \mathbf{r}$.
\end{enumerate}

The tensor products $\Qbound{i}{r} \otimes T_\mathbf{r}$ inside the embedding (Eq.~\eqref{eq:embedded-bound}) are direct products of the corresponding edge-induced subgraphs $G[\Qbound{i}{r}] \times H[T_\mathbf{r}]$. The sum over labels $\mathbf{r}$, which adds together adjacency matrices of the products modulo $2$, composes the edge sets of the corresponding graphs as a disjunctive union. In this way, the entire crystal complex may be constructed directly as a graph from a given unit cell and a lattice of arbitrary size.

\section{Characterization of noisy channels}\label{subsec:characterization_of_noisy_channels}
The general process in our simulations can be described as an $n$-qubit quantum circuit $C$ composed of Clifford operations, ending in a projective Pauli basis measurement $P^\fatm$ with outcomes $\fatm=\{m_1, m_2, \dots, m_l\}$ of (part) of the evolved state. For the sake of completeness, we also assume that the circuit operates on an ancillary input system $A$ with a stabilizer state $\ket{\psi_0}$. Such a circuit might represent an entanglement distillation circuit, operating on a mixed Bell pair $\rho$ and an ancillary Bell pair to distill it with. Alternatively, it may represent the action of a measurement-based fault-tolerant channel, where $\rho$ is the input code space, $\ket{\psi_0}$ is the state of all ancillary qubits in the channel, $C$ represents the C$Z$ gates of the cluster state and $P^\fatm$ is the final $X$-basis measurement of every ancillary qubit. The output code space is then (up to normalization) given by $\mathcal{E}_\fatm(\rho)$.

\begin{figure}[b]
    \centering
    \includegraphics[width=\linewidth]{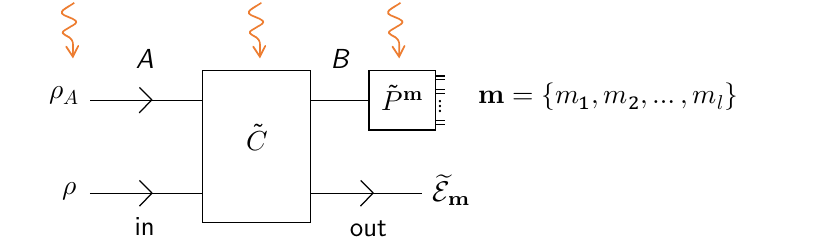}
    \caption[A noisy channel that can be characterized efficiently.]{A noisy channel that can be characterized efficiently, given that $\rho_A$ is a convex combination of stabilizer states, $\widetilde{C}$ is the ideal circuit $C$ with Pauli noise, and $\widetilde{P}^\fatm$ are ideal projectors followed by classical bit-flips of the outcome $\fatm$. Under suitable assumptions of the noise models, the $\widetilde{\mathcal{E}}_\fatm$ operation may be expressed as a mixture of ideal $\mathcal{E}_\fatm$ and Pauli operations.}
    \label{fig:noisy_channel}
\end{figure}

We consider three sources of noise, depicted schematically in Fig. \ref{fig:noisy_channel}. First of all, noisy ancillary input may not be a pure stabilizer state $\ket{\psi_0}$, but a mixture $\rho_A$ of possibly non-stabilizer states. Secondly, the circuit $C$ consists of imperfect operations, which we assume as ideal operations followed by a mixture of Pauli gates. Lastly, the projectors $P^\fatm$ may produce a ``wrong'' outcome $\widetilde{\fatm}$, which we model as a perfect operation $P^\fatm$ followed by classical bit-flips on $\fatm$.

Arbitrary noisy ancillary input states $\rho_A$ that differ from the noiseless input $\ket{\psi_0}$ cannot be simulated efficiently. If, on the other hand, $\rho_A$ may be approximated as a Pauli channel $\mathcal{P}$ acting on $\ket{\psi_0}$, the Pauli operators may be pushed through the circuit in the same way as Pauli noise coming from imperfect gates. The key idea is to pre-process $\rho_A$ by \emph{twirling}~\cite{Bennett1996b} with the stabilizers $s_k \in S_0$, where $S_0$ is the stabilizer group describing the state $\ket{\psi_0}$, \textit{i.e.}, to apply a trace-preserving channel $\mathcal{T}$ as
\begin{equation}
    \label{eq:twirl}
    \mathcal{T}\left(\rho_A\right) = \frac{1}{\abs{S_0}} \sum_{s_k\in S_0} s_k \rho_A s_k.
\end{equation}
In the context of Pauli twirling the input state, $\mathcal{T}$ corresponds to a Pauli channel $\mathcal{P}$ acting on the noiseless input state $\ket{\psi_0}$, with elements of a \emph{destabilizer} group $D_0$ of $S_0$ acting as the Pauli operators of the channel. A destabilizer group $D_0$ associated with a stabilizer group $S_0$ is a subgroup of the full Pauli group. It has the same size as $S_0$ and is generated with a set of operators that each anti-commute with a different generator of $S_0$ and commute with all other generators of $S_0$~\cite{aaronsonImprovedSimulationStabilizer2004}. A destabilizer group $D_0$ of $S_0$ can be used to decompose $\rho_A$ in a basis of states $\set{d_k\ket{\psi_0}}_{d_k\in D_0}$ and write
\begin{equation}
\rho_A=\sum_{d_m, d_p\in D_0} \lambda_{mp} d_m \ketbra{\psi_0}{\psi_0}d_p.
\end{equation}
We can now use this to write $\mathcal{T}(\rho_A)$ as
\begin{equation}
    \label{eq:twirl-as-pauli}
    \mathcal{T}\left(\rho_A\right) = \sum_{d_k\in D_0} p_k d_k\ketbra{\psi_0}{\psi_0}d_k = \mathcal{P}\left(\ketbra{\psi_0}{\psi_0}\right).
\end{equation}
Here, the prefactors $p_k$ are given by $p_k = \lambda_{kk} = \Braket{\psi_0|d_k \rho_A d_k|\psi_0}$. The trace $\Tr\left[\mathcal{T}\left(\rho_A\right)\right] = \sum_{d_k\in D_0} p_k = 1$ is preserved, such that prefactors $p_k$ that sum to unity may be interpreted as the probability of applying some $p_k$ in the Pauli channel $\mathcal{P}$. We see that twirling $\rho_A$ over the group $S_0$ removes all off-diagonal elements of the state in the $\set{d_k\ket{\psi_0}}_{d_k\in D_0}$ basis.

This shows how we may approximate noisy ancillary input state $\rho_A$ as a mixture of the ideal state $\ket{\psi_0}$ that is depolarized by a Pauli channel $\mathcal{P}$. In the same way, we may twirl non-Pauli noisy processes occurring during the application of the Clifford circuit $C$ to a Pauli form. Operators $P_i$ can now be propagated through the circuit $C$, forming another set of Pauli strings $P_j = C P_i C^\dagger$. Pauli operators $P_j$ now appearing after $C$ may each be split up into a string $P_j^{(B)}\equiv P'_j$ appearing on the ancillary $B$ system and a string $P_j^\text{(out)}\equiv P''_j$ appearing on the output system as $P_j \equiv P'_j \otimes P''_j$. The string $P'_j$ will commute with some projectors $P^{m_k} P'_j = P'_j P^{m_k}$ (where $m_k \in \fatm$), but anticommute with others as $P^{m_{k'}} P'_j = P'_j P^{\left(m_{k'} + 1\right)}$. The addition of two classical bits $a + b$ here is understood modulo 2, \textit{i.e.}, $a+1$ represents a bit-flip of $a$. We summarize both cases as a commutation relation $P^\fatm P'_j = P'_j P^{\fatm + \fatm_j}$, where $\fatm_j$ is a string of errors that represents the bit-flips due to the individual commutation relations above.

\begin{figure*}[t]
    \centering
    \includegraphics[width=0.85\textwidth]{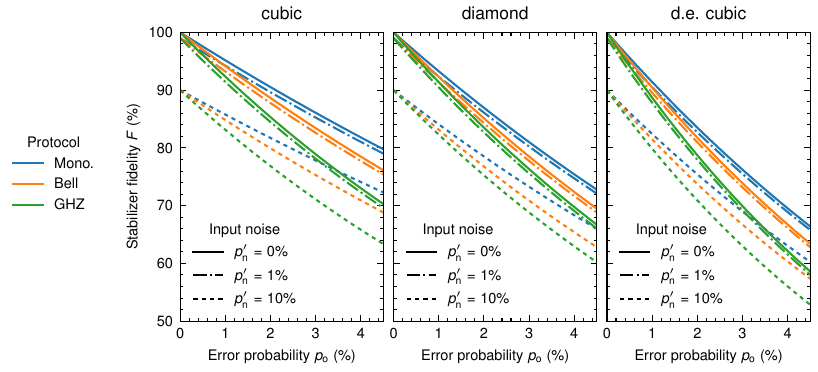}
    \caption[Stabilizer fidelities of monolithic and distributed protocols.]{Stabilizer fidelities of monolithic and distributed protocols for various architectures. The three protocols per architecture correspond to the three columns of Fig.~\ref{fig:distributed-architectures}, where the GHZ state is defined as weight-4 for cubic and double-edge cubic architectures and weight-3 for the diamond lattice, and a face with a Bell pair split is chosen for the diamond 7-node lattice. The noise probability $p_\textrm{o}$ describes both depolarizing gate noise with probability $p_\textrm{g}$ and faulty measurements with probability $p_\textrm{m}$. The quality of the network link is parameterized by $p'_\textrm{n}$ (see main text for details). Since we are only looking at the errors arising on a single face, the ordering of C$Z$ gates is not relevant for the stabilizer fidelity of that same face.}
    \label{fig:stabilizer-fidelities}
\end{figure*}

One can now derive an expression for the noisy operation $\widetilde{\mathcal{E}}_\fatm$ as a mixture of ideal operations with the addition of probabilistic Pauli strings:
\begin{equation}\label{eq:noisy-channel}
    \widetilde{\mathcal{E}}_\fatm (\rho) = \sum_j p_j P''_j \mathcal{E}_{\fatm + \fatm_j}(\rho) P''_j.
\end{equation}
The noisy channel $\widetilde{\mathcal{E}}_\fatm$ is a mixture of ideal operations $\mathcal{E}_{\fatm + \fatm_j}$ and Pauli noise, with the mixture arising due to classical bit-flips $\fatm_j$ that act on $\fatm$. Because we assumed each operation $\mathcal{E}_\fatm$ to be efficiently simulatable, the mixture may also be simulated efficiently.

On top of this, to account for faulty measurements, we assume that each projector $P^{m_k}$ has a fixed probability $p_\mathrm{m}$ of reporting the wrong outcome $\widetilde{m}_k \equiv m_k + 1$ (modulo 2), such that the noisy measurement channel $\widetilde{P}^{m_k}$ is given by a mixture
\begin{equation}
    \widetilde{P}^{m_k}(\rho) = \left(1-p_\mathrm{m}\right) P^{m_k} \rho\, P^{m_k} + p_\mathrm{m} P^{m_k+1} \rho\, P^{m_k+1}.
\end{equation}
For all measurement outcomes $\fatm=\{m_1, m_2, \dots, m_l\}$ this corresponds to the channel
\begin{equation}
\begin{split}
    \widetilde{P}^\fatm (\rho)&=\sum_f p_f P^{\fatm+\fatm_f}\rho\, P^{\fatm+\fatm_f}, \\
    p_f &\equiv (p_\mathrm{m})^{h(\fatm_f)}(1-p_\mathrm{m})^{l-h(\fatm_f)},
    \end{split}
\end{equation}
where $h(\fatm_f)$ is the \emph{Hamming weight} of the binary string $\fatm_f$ with measurement errors. 
The additional mixing of measurement outcomes does not change the form of the noisy channel as in Eq.~\eqref{eq:noisy-channel} but introduces additional terms $\mathcal{E}_{\fatm + \fatm_j + \fatm_f}$ with prefactors that reflect the probability of applying a particular configuration of measurement errors:
\begin{equation}
    \widetilde{\mathcal{E}}_\fatm (\rho) = \sum_{j,f} p_{j} p_{f} P''_j \mathcal{E}_{\fatm+\fatm_j+\fatm_f} (\rho) P''_j.
\end{equation}

\section{Statistical error in threshold values}\label{app:statistical_error}
To fit the thresholds, we assume a second-order polynomial model of the logical error probability $p_\mathrm{L}$ around the threshold crossing $p_\textrm{th}$ of the form~\citep{wangConfinementHiggsTransitionDisordered2003}
\begin{equation}
    p_\mathrm{L}(p,L) = p_{\mathrm{L},\textrm{th}} + c_1\left(p - p_{\textrm{th}}\right)L^{1/\nu} + c_2\left(p - p_{\textrm{th}}\right)^2 L^{2/\nu}, 
\end{equation}
where $L$ corresponds to the lattice size used with the error probability $p$, and $p_{\mathrm{L},\textrm{th}}$, $p_\mathrm{th}$, $c_1$, $c_2$, and $\nu$ are fitting parameters. 
The fit is obtained through non-linear least-squares minimization of the residuals $R_i = \left(p_{\mathrm{L}}(p,L) - \hat{p}_{i}\right)/\hat{\sigma}_{i}$ for every data point $i$ with the observed logical error rate $\hat{p}_{i}$ and the associated standard deviation $\hat{\sigma}_{i}$. 
Confidence intervals for threshold crossings $p_\textrm{th}$ are taken from the standard errors of the least-squares approximation.

In the results of Sec.~\ref{sec:results}, error bars of the logical error rate are given as $95\%$ confidence intervals. Sometimes, they are too small to be discernible. The threshold value is highlighted with a $95\%$ confidence interval based on its least-squares estimate of a second-order polynomial of the logical error rate around the threshold value.

\section{Stabilizer fidelities for distributed setting}\label{sec:stabilizer_fidelities_for_distributed_setting}
In this appendix, we gauge the fidelity of the stabilizer operator supported by each face, for several face splittings in the distributed setting. 
For these results, we model the entangled states as Bell and GHZ states that are depolarized to a diagonal form. For these states, the primary component $S_0 \equiv \Braket{X_0\dots X_{n-1}, Z_0Z_1, \dots, Z_0Z_{n-1}}$ has probability $1-p'_\textrm{n}$ and all off-diagonal terms $\Braket{\pm X_0\dots X_{n-1}, \pm Z_0Z_1, \dots, \pm Z_0Z_{n-1}}$ with at least one negative sign are uniformly distributed with probabilities $p'_\textrm{n}/(2^n-1)$. With $n=2$, this state corresponds to the Werner state of Eq.~\eqref{eq:werner_state}. Furthermore, just as in Sec.~\ref{subsec:distributed_thresholds}, we assume that state preparation and measurements invert the state with probability $p_\textrm{m}\equiv p_\textrm{o}$, and that C$Z$ gates are followed by a depolarizing channel with probability $p_\textrm{g}\equiv p_\textrm{o}$. 

For each of the geometries, the stabilizer fidelity of one of the faces is calculated under this circuit-based noise model. Fidelities are calculated as the overlap of the resulting mixed state with the ideal cluster state stabilizer. A strong simulation for each of the circuits provides an exact closed-form expression of the fidelity. We show these results for each of the lattices and the three different protocols (monolithic, Bell, and GHZ) corresponding to the three columns of Fig.~\ref{fig:stabilizer-fidelities}.
Based on these results, we see that an increase in the number of splittings produces worse stabilizer fidelities. These differences are more pronounced for the 4-partite GHZ states when compared to the 3-partite GHZ state in diamond. Distributed architectures are likely unable to compete with monolithic protocols, unless we combine entanglement generation protocols with better-quality GHZ states through entanglement distillation.

\end{document}